\documentclass[a4paper,fleqn,usenatbib]{mnras}
\usepackage[T1]{fontenc}
\usepackage{ae,aecompl}
\usepackage{amsmath}
\usepackage{amsfonts}
\usepackage{amssymb}
\usepackage{graphicx}
\usepackage{color,xcolor}
\usepackage{tabularx}
\newcolumntype{L}{>{\raggedright\arraybackslash}X}
\def\mnras{MNRAS}
\def\aap{A$\&$A}
\def\nat{Nature}
\def\apj{ApJ}
\def\apjl{ApJ Letters}
\def\apjs{ApJS}

\def\aj{AJ}

\newcommand{\msun}{\, {\rm M}_{\odot}}
\newcommand{\msunyr}{\, {\rm M}_{\odot}\,{\rm yr^{-1}}}

\newcommand{\au}{\, {\rm au}}

\title[Viscosity or MHD winds?]{Photoevaporation obfuscates the distinction between wind and viscous angular momentum transport in protoplanetary discs}
\author[G. A. L. Coleman et al]{Gavin A. L. Coleman\thanks{Email: gavin.coleman@qmul.ac.uk}, Joseph K. Mroueh, Thomas J. Haworth\\
Astronomy Unit, Department of Physics and Astronomy, Queen Mary University of London, Mile End Road, London, E1 4NS, UK}
\date{Accepted 2023 November 27; Received 2023 November 27; in original form 2023 September 25}
\pubyear{2024}
\begin{document}
\label{firstpage}
\pagerange{\pageref{firstpage}--\pageref{lastpage}}
\maketitle
\begin{abstract}
How protoplanetary discs evolve remains an unanswered question. Competing theories of viscosity and magnetohydrodynamic disc winds have been put forward as the drivers of angular momentum transport in protoplanetary discs. These two models predict distinct differences in the disc mass, radius and accretion rates over time, that could be used to distinguish them. However that expectation is built on models that do not include another important process - photoevaporation, both internally by the host star and externally by neighbouring stars. In this work we produce numerical models of protoplanetary discs including viscosity, magnetohydrodynamic disc winds, and internal and external photoevaporation. We find that even weak levels of external photoevaporation can significantly affect the evolution of protoplanetary discs, influencing the observable features such as disc radii, that might otherwise distinguish between viscous and wind driven discs. Including internal photoevaporation further suppresses differences in evolution between viscous and wind driven discs. This makes it much more difficult than previously anticipated, to use observations of nearby star forming regions to determine whether discs are viscous or wind driven. Interestingly we find that evolved protoplanetary discs in intermediate FUV environments may be the best cases for differentiating whether they evolve through viscosity or magnetohydrodynamic disc winds. Ultimately this work demonstrates the importance of understanding what are the key evolutionary processes and including as many of those as possible when exploring the evolution of protoplanetary discs. 
\end{abstract}
\begin{keywords}
 accretion, accretion discs -- protoplanetary discs -- (stars:) circumstellar matter
\end{keywords}

\section{Introduction}
\label{sec:intro}

Planet formation is likely to be intrinsically linked to the evolution of the circumstellar discs of material around young stars \citep[e.g.][]{2018ApJ...869L..41A, 2018A&A...617A..44K, 2018ApJ...860L..13P, 2019Natur.574..378T}. However, understanding this is an extremely challenging problem. We observe each disc at only a snapshot in its evolution, and any given disc is difficult to age. This is further complicated by the fact that star formation happens over time in stellar clusters \citep[e.g.][]{2014ApJ...787..108G, 2014ApJ...787..109G, Qiao22, 2023MNRAS.520.6159C}. Since discs have a huge radial variation in composition, temperature, density and optical depth, any single observation also only probes a limited region of the disc \citep[see e.g. Figure 1 of][]{2016PASA...33...59S}. 

Whilst there are numerous observational challenges, there are many processes that are thought to affect how and on what time-scales protoplanetary discs evolve. For example discs evolve through accreting on to the central star \citep[e.g.][]{1981ARA&A..19..137P, 2016A&A...591L...3M, 2020A&A...639A..58M, 2021A&A...650A.196M}, internal photoevaporative \citep[e.g.][]{Alexander07, Owen12, 2017RSOS....470114E} or magnetically driven winds \citep{BalbusHawley1991, 2007prpl.conf..277P, Suzuki09, 2014prpl.conf..411T, 2016ApJ...818..152B, Tabone22, 2023ASPC..534..567P} as well as winds driven by external irradiation from nearby massive stars \citep[e.g.][]{1994ApJ...436..194O, 2000ApJ...539..258R, 2004ApJ...611..360A, 2005AJ....130.1763S, 2016ApJ...826L..15K,  2019MNRAS.485.3895H, 2021MNRAS.501.3502H}.
Typically most disc evolution models only include one or two of these processes.

Focusing  on accretion through the disc and onto the central star, the main theories accounting for the necessary angular momentum transport are through viscous accretion or magnetohydrodynamic (MHD) driven disc winds. Traditionally, turbulence in the disc, driven by the magnetorotational instability (MRI)\citep{BalbusHawley1991} was thought to drive accretion. This turbulence was modelled as a form of viscosity \citep{Shak,Lynden-BellPringle1974}, where shearing rings of gas would drive turbulent transport through the disc. Using the popular $\alpha$-disc model \citep{Shak}, this was found to require $\alpha\sim 10^{-3}$--$10^{-2}$ in order to match observed accretion rates \citep{King07,Rafikov17}. More recently, numerous works have provided additional observational estimates for the viscosity parameter $\alpha$, through for example, accretion on to the central star, or the vertical extent of dust discs \citep{Flaherty17,Ansdell18,Trapman20,Villenave20,Villenave22}.
They typically find a range in $\alpha$ from $\sim10^{-4}$ to $\sim 10^{-2}$ with a median between $3\times10^{-4}$--$3\times10^{-3}$ \citep[see][for a recent review]{Rosotti23}.

On the other hand, it has also been widely accepted that large areas of protoplanetary discs are too dense to be sufficiently ionised and thus coupled the stellar magnetic field \citep{Gammie96}. For such regimes, where the gas is only partially ionised, non-ideal MHD effects dominate and suppress the MRI \citep{Armitage11}, resulting in a non-zero magnetic flux that then drives magnetised disc winds \citep[e.g.][]{Suzuki09,Bai13,Fromang13,Gressel15}. The properties of such winds would then be determined by the properties of the magnetic field, rather than the local disc conditions \citep{Lesur23}. As the winds then transport mass and angular momentum away from the disc, this then drives an accretion flow to conserve angular momentum \citep{Salmeron11}. Recently, analytic disc models have been formulated that include MHD disc winds as the driver of accretion through the disc, finding qualitative differences between their evolution and that of purely viscous discs \citep{Tabone22}.

Whilst we cannot directly measure whether discs are viscous or MHD wind driven, we can: place constraints on the viscous $\alpha$ parameter which indirectly constrains the possible driving mechanism \citep[e.g.][]{Rosotti23}, find evidence for magnetised disc winds \citep[e.g.][]{2021ApJS..257...16B, 2023ApJ...944...63L,Launhardt23}, or study populations of discs at different evolutionary stages to compare with the predictions of work such as \cite{Tabone22,Somigliana23}. A factor that may complicate the ability to unambiguously distinguish MHD and viscously driven evolution is photoevaporation, where high energy radiation from either the central star \citep[internal photoevaporation, e.g. ][]{2022EPJP..137.1357E}, or from nearby stars \citep[external photoevaporation, e.g.][]{2022EPJP..137.1132W} heat the outer layers of protoplanetary discs and drive a thermal wind. Internal photoevaporative winds and MHD disc winds originate from similar portions of the disc, which can make them difficult to distinguish. Both internal and external photoevaporative winds can affect the disc mass, radius, lifetime and accretion properties over time \citep[e.g.][]{2018MNRAS.477.4131J, 2020MNRAS.498.2845S, Coleman22, Ercolano23}. 

With multiple processes acting on protoplanetary discs, previous works have explored differences between individual factors. For example \citet{Tabone22} analytically compared purely viscous discs to purely wind driven discs, finding qualitative differences in disc radii and mass accretion rates. \cite{2020MNRAS.497L..40W} studied the interplay between viscous resupply of an externally driven wind, finding that the rate of spreading moderates the external mass loss rate in older systems. More recently \citet{Alexander23} compared analytic solutions for wind driven discs and viscous discs, including internal photoevaporation. They concentrated on mass accretion rates, finding that sufficiently large observational datasets could differentiate between the models. Exploring the effects of photoevaporation on viscous discs, \citet{Coleman22} found five different evolution pathways for protoplanetary discs, where observations could possibly constrain the relative importance of internal and external photoevaporation for any given disc.

In this work, we systematically include all of these processes in a single numerical model. We adapt the analytic model found in \citet{Tabone22} and couple to the prescribed mass loss rates for internal photoevaporation \citep[following][]{Picogna21}, and interpolate within the newly updated \textsc{fried} grid of mass loss rates due to external photoevaporation \citep{Haworth23}. Our aim is to determine if and what differences there are in the evolution of viscously and wind-driven protoplanetary discs in different star forming environments. These differences could then be compared to future observations, hopefully placing further constraints on the processes that are at play, and to what their properties are, within protoplanetary discs. 

This paper is organised as follows.
Section \ref{sec:physical_model} outlines the disc evolution and photoevaporation models as well as the simulation parameters.
In sect. \ref{sec:results} we explore the effects of the local star forming environment on the evolution of viscous or wind driven discs.
We include the effects of internal photoevaporation and examine its consequences in sect. \ref{sec:internal_pe}.
Section \ref{sec:discussion} discusses possible observations that could differentiate between viscous and MHD wind driven models.
Finally we draw our conclusions in sect. \ref{sec:conclusions}.

\section{Physical Model and Parameters}
\label{sec:physical_model}

Protoplanetary discs lose mass by accretion onto the central star and through photoevaporative winds launched from the disc surface layers.
To account for these processes we follow \citet{Tabone22} and include both contributions from a traditional 1D viscous $\alpha$ disc model \citep{Shak}, in addition to contributions from an MHD disc wind.
The MHD disc wind is controlled by two parameters: $\alpha_{\rm DW}$, which quantifies the angular momentum extracted by the wind, and $\lambda$, the magnetic lever arm parameter which determines the amount of mass lost in the wind.
The gas surface density $\Sigma$ at radius $r$, is therefore evolved solving an updated diffusion equation
\begin{equation}
\label{eq:diffusion}
\begin{split}
    \dot{\Sigma}(r)=&\dfrac{1}{r}\dfrac{d}{dr}\left[3r^{1/2}\dfrac{d}{dr}\left(\nu\Sigma r^{1/2}\right)\right]+\dfrac{3}{2r}\dfrac{d}{dr}\left[\dfrac{\alpha_{\rm DW}\Sigma c_{\rm s}^2}{\Omega}\right]\\
    &-\dfrac{3\alpha_{\rm DW}\Sigma c_{\rm s}^2}{4(\lambda-1)r^2\Omega}-\dot{\Sigma}_{\rm PE}(r)
\end{split}
\end{equation}
where $\nu=\alpha_{\rm v} H^2\Omega$ is the disc viscosity with viscous parameter $\alpha_{\rm v}$ \citep{Shak}, $H$ is the disc scale height and $\Omega$ the Keplerian frequency.
The second and third components on the right hand side of eq. \ref{eq:diffusion} represent the change in surface density due to the angular momentum extracted by the wind, and the the mass lost in the MHD wind itself.
The fourth term represents mass extracted by photoevaporative winds, which we discuss below.
Note there are two components for $\alpha$ in eq. \ref{eq:diffusion}, those being the measure of turbulence for viscosity $\alpha_{\rm v}$, and the measure of angular momentum extracted in the wind $\alpha_{\rm DW}$. The equation does not however specify the relationship between the two $\alpha$ parameters. Therefore we follow \citet{Tabone22} and define
\begin{equation}
    \alpha = \alpha_{\rm v} + \alpha_{\rm DW}
\end{equation}
where $\alpha$ quantifies the total torque exerted by turbulence and the MHD disc wind. Further, we define
\begin{equation}
    \psi = \dfrac{\alpha_{\rm DW}}{\alpha_{\rm v}}
\end{equation}
as a parameter that quantifies the relative strength between the two $\alpha$ parameters. In this work we explore between $\psi=10^{-4}$ as a viscous case, to $\psi = 10^4$ as a wind driven case.

Our models also include depletion of the disc through photoevaporative winds.
This is expressed by $\dot{\Sigma}_{\rm PE}(r)$ in eq. \ref{eq:diffusion}.
Following \citet{Coleman22} we include internal photoevaporative winds due to X-ray photons coming from the central star (detailed in section \ref{sec:internalPhoto}) as well as winds launched from the outer disc by far ultraviolet (FUV) radiation emanating from nearby massive stars (e.g. O-type stars, see section \ref{sec:externalPhoto}).
When we utilise both internal and external photoevaporation regimes, we assume that the photoevaporative mass loss rate at any radius in the disc is the maximum of the internally and externally driven rates 
\begin{equation}
    \dot{\Sigma}_{\rm PE}(r) ={\rm max}\left(\dot{\Sigma}_{\rm I,X}(r),\dot{\Sigma}_{\rm E,FUV}(r)\right)
\end{equation}
where the subscripts I and E refer to contributions from internal and external photoevaporation.

We assume that the disc is in thermal equilibrium, where the temperature is calculated by balancing irradiation heating from the central star, background heating from the residual molecular cloud, viscous heating and blackbody cooling
To attain this equilibrium, we follow \citet{Coleman21} and use an iterative method to solve the following equation \citet{Dangelo12}
\begin{equation}
    Q_{\rm irr} + Q_{\nu} + Q_{\rm cloud} - Q_{\rm cool} = 0
\end{equation}
where $Q_{\rm irr}$ is the radiative heating rates due to the central star, $Q_{\nu}$ is the viscous heating rate per unit area of the disc, $Q_{\rm cloud}$ is the radiative heating due to the residual molecular cloud, and $Q_{\rm cool}$ is the radiative cooling rate.

\subsection{Internal Photoevaporation}
\label{sec:internalPhoto}
The absorption of high energy radiation from the host star by the disc can heat the gas above the local escape velocity, and hence drive internal photoevaporative winds. EUV irradiation creates a layer of ionised hydrogen with temperature $\sim$10$^4$~K \citep{Clarke2001}, however X-rays penetrate deeper into the disc and are still capable of heating up to around $\sim$10$^4$~K \citep{Owen10} so for low mass stars are expected to generally dominate over the EUV for setting the mass loss rate. FUV radiation penetrates deeper still, creating a neutral layer of dissociated hydrogen with temperature of roughly 1000K \citep{Matsuyama03}. The overall interplay between the EUV, FUV and X-rays is a matter of ongoing debate.  \cite{Owen12} find that including the FUV heating simply causes the flow beneath the sonic surface to adjust, but otherwise retains the same mass loss rate. However others suggest a more dominant role of the FUV \citep[e.g.][]{2009ApJ...705.1237G,2015ApJ...804...29G}. Recent models including all three fields suggest a more complicated interplay \citep[e.g.][]{2017ApJ...847...11W, 2018ApJ...865...75N,Ercolano21}.  The outcome also depends sensitively on how the irradiated spectrum is treated \citep{2022MNRAS.514..535S}. 

The radiation hydrodynamic models of \cite{Owen12} used pre-computed X-ray driven temperatures as a function of the ionisation parameter ($\xi = L_X / n /r^2$) wherever the column to the central star is less than $10^{22}$cm$^{-2}$ (and hence optically thin). This approach has since been updated with a series of column-dependent temperature prescriptions \citep{Picogna19,Ercolano21,Picogna21}.  

We follow \citet{Picogna21} who further build on the work of \cite{Picogna19} and \cite{Ercolano21} and find that the mass loss profile from internal X-ray irradiation is approximated by
\begin{equation}
\label{eq:sig_dot_xray}
\begin{split}
\dot{\Sigma}_{\rm I,X}(r)=&\ln{(10)}\left(\dfrac{6a\ln(r)^5}{r\ln(10)^6}+\dfrac{5b\ln(r)^4}{r\ln(10)^5}+\dfrac{4c\ln(r)^3}{r\ln(10)^4}\right.\\
&\left.+\dfrac{3d\ln(r)^2}{r\ln(10)^3}+\dfrac{2e\ln(r)}{r\ln(10)^2}+\dfrac{f}{r\ln(10)}\right)\\
&\times\dfrac{\dot{M}_{\rm X}(r)}{2\pi r} \dfrac{\msun}{\au^2 {\rm yr}}
\end{split}
\end{equation}
where
\begin{equation}
\label{eq:m_dot_r_xray}
    \dfrac{\dot{M}_{\rm X}(r)}{\dot{M}_{\rm X}(L_{X})} = 10^{a\log r^6+b\log r^5+c\log r^4+d\log r^3+e\log r^2+f\log r+g}
\end{equation}
where $a=-0.6344$, $b=6.3587$, $c=-26.1445$, $d=56.4477$, $e=-67.7403$, $f=43.9212$, and $g=-13.2316$.
We follow \citet{Komaki23} and apply a simple approximation to the outer regions of the disc where the internal photoevaporation rates drop to zero. The reasons given for this sudden drop is that the wind itself blocks radiation from heating the outer regions of protoplanetary discs. However these do not take into account the effects of when the disc and/or the wind become optically thin and therefore ineffective at blocking the radiation.
The temperature of X-ray irradiated gas varies from $\sim 10^3$--$10^4$ K depending on the distance in the disc \citep[e.g.][]{Owen10}. To be conservative we define the radius at which the internal photoevaporation scheme drops off as the gravitational radius for $10^3$ K gas.
We therefore apply the following approximation at radial distances greater than $r_{\rm rgx}$
\begin{equation}
    \dot{\Sigma}_{\rm I,X,ap} = 37.86\times\dot{\Sigma}_{\rm rgx}\left(\frac{r}{r_{\rm rgx}}\right)^{-1.578}
\end{equation}
where $\dot{\Sigma}_{\rm rgx}$ is equal to eq. \ref{eq:sig_dot_xray} calculated at $r=r_{\rm rgx}$, and
\begin{equation}
    r_{\rm rgx} = \dfrac{GM_*}{c_{\rm s}^2}
\end{equation}
where $c_{\rm s}$ is the sound speed for gas of temperature $T=10^3 K$, and $\mu=2.35$.
In the outer regions of the disc the loss in gas surface density due to internal photoevaporation then becomes
\begin{equation}
    \dot{\Sigma}_{\rm I}(r) = \max(\dot{\Sigma}_{\rm I,X}(r),\dot{\Sigma}_{\rm I,X,ap})
\end{equation}

Following \cite{Ercolano21} the integrated mass-loss rate, dependant on the stellar X-ray luminosity, is given as
\begin{equation}
    \log_{10}\left[\dot{M}_{X}(L_X)\right] = A_{\rm L}\exp\left[\dfrac{(\ln(\log_{10}(L_X))-B_{\rm L})^2}{C_{\rm L}}\right]+D_{\rm L},
\end{equation}
in $\msunyr$, with $A_{\rm L} = -1.947\times10^{17}$, $B_{\rm L} = -1.572\times10^{-4}$, $C_{\rm L} = -0.2866$, and $D_{\rm L} = -6.694$.

\subsection{External Photoevaporation}
\label{sec:externalPhoto}

In addition to internal winds driven by irradiation from the host star, winds can also be driven from the outer regions of discs by irradiation from external sources. Massive stars dominate the production of UV photons in stellar clusters and hence dominate the external photoevaporation of discs. External photoevaporation has been shown to play an important role in setting the evolutionary pathway of protoplanetary discs \citep{Coleman22}, their masses \citep{2014ApJ...784...82M, 2017AJ....153..240A, 2023A&A...673L...2V}, radii \citep{2018ApJ...860...77E} and lifetimes \citep{2016arXiv160501773G, 2019MNRAS.490.5678C, 2020MNRAS.492.1279S, 2020MNRAS.491..903W} even in weak UV environments \citep{2017MNRAS.468L.108H}.
We do not include shielding of the protoplanetary discs, i.e. by the nascent molecular cloud, that has been shown to have an effect on the effectiveness of external photoevaporation \citep{Qiao22,Qiao23, 2023MNRAS.520.5331W}, but instead will infer it's effects by examining weaker environments.

In our simulations, the mass loss rate due to external photoevaporation is calculated via interpolating over the recently updated \textsc{fried} grid \citep{Haworth23}.
This new grid expands on the original version of \textsc{fried} \citep{2018MNRAS.481..452H} in terms of the breadth of parameter space in UV field, stellar mass, disc mass and disc radius. The new grid also provides the option to use different PAH-to-dust ratios, which is important because polycyclic aromatic hydrocarbons (PAH) can provide the main heating mechanism in a photodissociation region (PDR, which is the region at the base of an external photoevaporative wind) and their abundance is uncertain. Finally, the new grid provides the option to control whether or not grain growth has taken place in the disc, which affects the opacity in the wind since only small grains are entrained.

The \textsc{fried} grid provides mass loss rates for discs irradiated by FUV radiation as a function of the star/disc/FUV parameters. In our simulations, we determine the mass loss rate at each time step by linearly interpolating \textsc{fried} in three dimensions: disc size $R_{d}$, disc outer edge surface density $\Sigma_{\textrm{out}}$ and FUV field strength $F_{\rm{FUV}}$.

We evaluate the \textsc{fried} mass loss rate at each radius from the outer edge of the disc down to the radius that contains 80$\%$ of the disc mass. We choose this value as 2D hydrodynamical models show that the vast majority of the mass loss from external photoevaporation, comes from the outer 20\% of the disc \citep{2019MNRAS.485.3895H}. The change in gas surface density is then calculated as
\begin{equation}
    \dot{\Sigma}_{\textrm{ext, FUV}}(r) = G_{\rm sm} \frac{\dot{M}_\textrm{{ext}}(R_{\textrm{\textrm{max}}})}{\pi(R^2_\textrm{{d}} - {R_{\textrm{\textrm{max}}}}^2)+A_{\rm sm}}, 
\end{equation}
where $A_{\rm sm}$ is a smoothing area equal to 
\begin{equation}
A_{\rm sm} = \dfrac{\pi(R_{\rm max}^{22}-(0.1 R_{\rm max})^{22})}{11R_{\rm max}^{20}}
\end{equation}
and $G_{\rm sm}$ is a smoothing function
\begin{equation}
    G_{\rm sm} = \dfrac{r^{20}}{R_{\rm max}^{20}}.
\end{equation}

The {\sc fried} grid contains multiple subgrids that vary the PAH-to-dust ratio ($\rm f_{PAH}$) and specify whether or not grain growth has occurred.
The effects of using different combinations of these parameters will be explored in future work, but we do not expect such changes to affect the differences between viscous and MHD wind driven discs.
The combination we use here is $\rm f_{PAH}=1$ (an interstellar medium, ISM,-like PAH-to-dust ratio) and assume that grain growth has occurred in the outer disc, depleting it and the wind of small grains which reduces the extinction in the wind and increases the mass loss rate compared to when dust is still ISM-like. This combination of parameters results in PAH-to-gas abundances comparable to our limited observational constraints on that value \citep{2013ApJ...765L..38V}. 

\subsection{Simulation Parameters}

Whilst our previous work examined the evolution of viscously evolving discs, including internal and external photoevaporation, around stars of different masses \citep{Coleman22}, here for simplicity, we only consider Solar mass stars. In future work, we will explore populations of stars with varying masses and in varying environments.
To explore the differences that arise from viscous or MHD-wind driven discs, we vary the ratio of the MHD-wind to turbulent viscous transport, $\psi$, in log intervals of 0.1 between $\psi = 10^{-4}$, that being the viscosity dominated case, to $\psi = 10^4$, that being the MHD-wind dominated case. Note that $\psi=1$ represents the hybrid case, where $\alpha$ is evenly split between viscosity and the MHD wind.
Numerous works have provided observational estimates for the viscosity parameter $\alpha$ \citep{Isella09,Andrews10,Pinte16,Flaherty17,Trapman20,Villenave20,Villenave22}.
To ensure consistency, we assume that the maximum that our value for $\alpha_{\rm v}$ can be is $10^{-3}$, consistent with estimates \citep[see][for a recent review]{Rosotti23}.
This is also consistent with our previous disc evolution scenarios \citep{Coleman22}, as well as planet formation scenarios that show that $\alpha\le 10^{-3}$ in order to form circumbinary systems similar to BEBOP-1 \citep{Standing23,Coleman24}.
For the external photoevaporative mass loss rates, we vary the strength of the local environment, ranging from 10 $\rm G_0$ to $10^{5} \rm G_0$, which spans most of the range found in star forming regions \citep{2008ApJ...675.1361F, 2020MNRAS.491..903W}.
X-ray luminosities are observed to vary by up to two orders of magnitude even for stars of the same mass, due to a combination of measurement uncertainty and genuine intrinsic differences in X-ray activity levels, which are time varying \citep[see figure 1 of][and the associated discussion]{Flaischlen21}.
Whilst we do not perform a parameter study including variations in X-ray luminosities, we take $L_X = 10^{30} \rm erg s^{-1}$ to account for the central star driving an internal photoevaporative wind.
We define this as a weak photoevaporative wind, since the X-ray luminosity chosen is half a magnitude smaller than the average found for Solar mass stars in nearby clusters \citep[$L_{\rm X}=10^{30.5} \rm erg s^{-1}$;][]{Flaischlen21}. With this X-ray luminosity, it yields an integrated mass loss rate of $5.4\times10^{-8} \msunyr$, but this value would be the maximum possible total internal photoevaporative mass loss rate, since we assume the mass loss due to photoevaporation at any given point in the disc is the maximum of the internal and external photoevaporative winds. Note that any plausible value for $L_{\rm X}$ within these simulations would give mass loss rates due to internal photoevaporation comparable to or exceeding the stellar accretion rate, as this is a natural consequence of internal photoevaporation \citep{Ercolano09,Owen10,Picogna19,Ercolano21}.

\begin{figure*}
\centering
\includegraphics[scale=0.5]{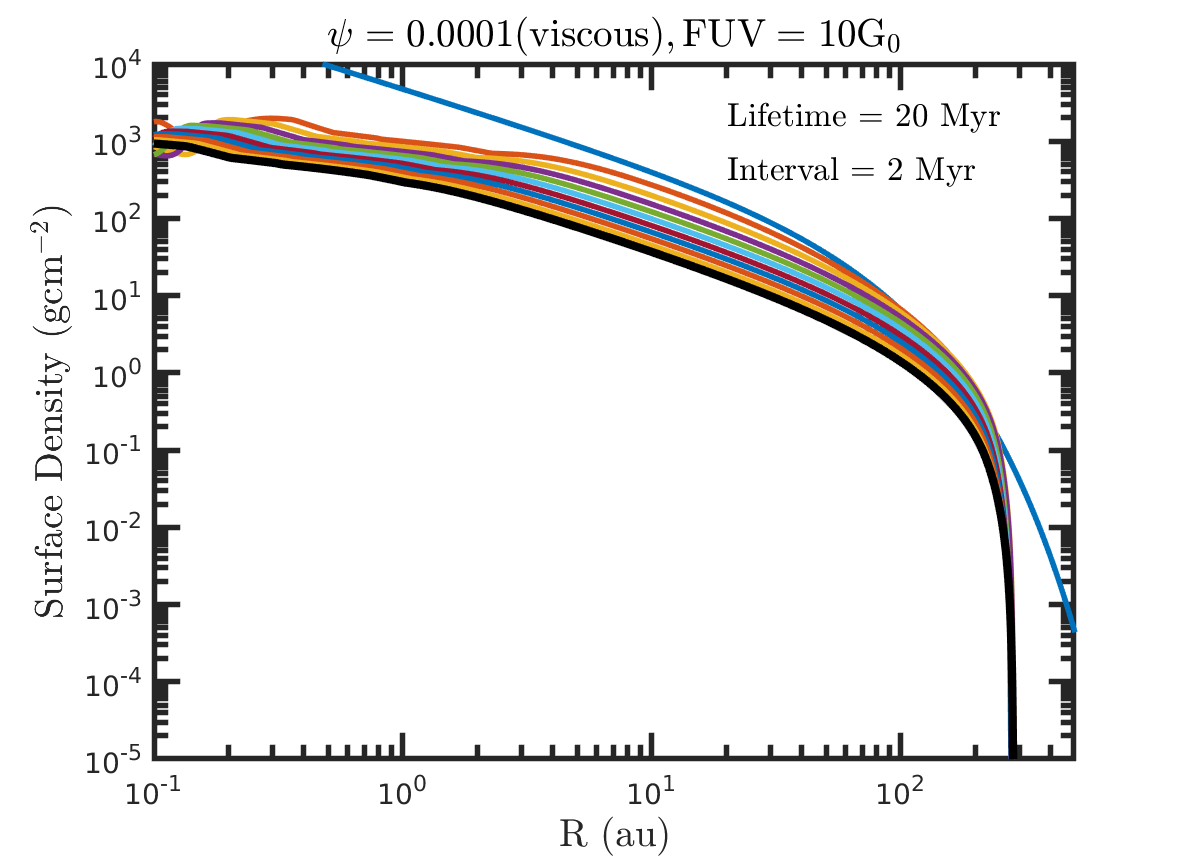}
\includegraphics[scale=0.5]{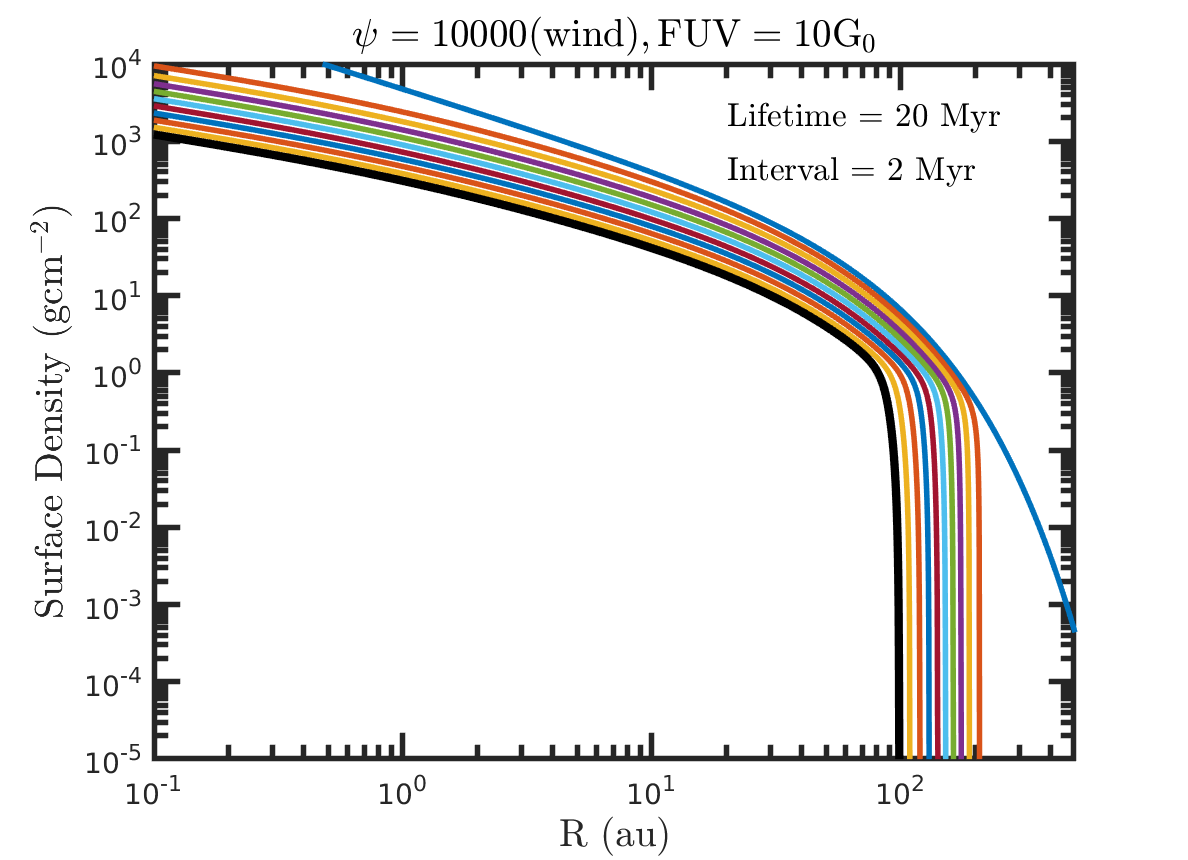}
\caption{Temporal evolution of gas surface densities for viscous discs (left panel), and wind-dominated discs (right panel). The uppermost blue line is the initial profile, whilst the black line shows the final output. The interval between profiles here is equal to 2 Myr.}
\label{fig:surface_density}
\end{figure*}

We initialise our discs following \citet{Lynden-BellPringle1974}
\begin{equation}
    \Sigma = \Sigma_0\left(\frac{r}{R_{\rm C}}\right)^{-1}\exp{\left(-\frac{R}{R_{\rm C}}\right)}
\end{equation}
where $\Sigma_0$ is the normalisation constant set by the total disc mass, (for a given $R_{\rm C}$), and $R_{\rm C}$ is the scale radius, which sets the initial
disc size, taken here to be equal to 50 $\au$.
For the initial mass of the disc we follow \citet{Haworth20} where from hydrodynamic simulations they find the maximum disc mass $M_{\rm d, max}$ that a gas disc of radius $r_{\rm ini}$, and a radial slope of -1 around a star of mass $M_*$ can be before becoming gravitationally unstable 
\begin{equation}
    \label{eq:max_disc_mass}
    \dfrac{M_{\rm d, max}}{M_*} = 0.17 \left(\dfrac{r_{\rm ini}}{100\au}\right)^{1/2}\left(\dfrac{M_*}{\msun}\right)^{-1/2}.
\end{equation}
Given that we initialise the disc with the \citet{Lynden-BellPringle1974} self similar solution, and not a constant slope, we take $r_{\rm ini} = 100\au$, twice that of the scale radius, in order to obtain appropriate protoplanetary disc masses that are gravitationally stable. Therefore in this work, with $r_{\rm ini} = 100\au$, the initial disc mass corresponds to $M_{\rm d}=0.17\msun$.
Table \ref{tab:parameters} shows the simulation parameters that we used in this work.

\begin{table}
    \centering
    \begin{tabular}{c|c}
    \hline
    Parameter & Value \\
    \hline
        $r_{\rm ini} (\au)$ & 100 \\
        $M_{\rm d,max} (\msun)$ & 0.17 \\
        $R_{\rm C} (\au)$ & 50 \\
        $r_{\rm in} (\au)$ & 0.1 \\
        $r_{\rm out} (\au)$ & 500 \\
        $\log_{10}(L_X)$ & 30 \\
        $\alpha$ & $10^{-3}$ \\
        $\psi$ & $10^{-4}$--$10^4$ \\
        $\rm G_0$ & $10^1$--$10^5$ \\
        $f_{\rm pah}$ & 1 \\
    \hline
    \end{tabular}
    \caption{Simulation Parameters. They are, from top to bottom, the initial disc radius, the maximum stable initial disc mass (we use half of this value in our initial conditions), the X-ray luminosity, the viscous $\alpha$ parameter and the external FUV radiation field strength and the assumed PAH-to-dust ratio relative to the ISM. }
    \label{tab:parameters}
\end{table}

\section{Effect of the local radiation Environment}
\label{sec:results}

The main focus of this section is on the effects of the local radiation environment, through external photoevaporative winds, on the evolution of discs that otherwise evolve through either viscosity or MHD-winds.
Within the simulation parameters we use, this corresponds to $\psi=10^{-4}$ for the viscous disc, and $\psi=10^4$ for the wind-driven disc.
These are most equivalent to the $\psi=0$ and $\psi=\infty$ in \citet{Tabone22}.
Initially we will explore the effects of a weak FUV environment, equivalent to 10$\rm G_0$, and similar to the FUV field strength found in star forming regions such as Taurus and Lupus \citep[e.g.][]{2016ApJ...832..110C}.
For this section, we do not include the effects of internal photoevaporation, so that we can systematically determine the effects of the local environment on the evolution of viscous or MHD wind driven discs.

In fig. \ref{fig:surface_density} we show the temporal evolution of the gas surface densities for a viscous dominated disc (left panel), and a wind dominated disc (right panel).
Because of the (weak) external FUV field, both discs evolve in a similar way at the start of their lifetimes, where they begin to truncate and reduce in size.
For the viscous case, the disc truncates to $\sim300\au$ where the viscous expansion rate equals the mass loss rate through external photoevaporation \citep{2007MNRAS.376.1350C}.
Over the rest of the disc lifetime here, the disc continues to accrete onto the central star and lose mass through the external photoevaporative wind.
In contrast, for the wind dominated disc (where there is no outwards expansion like in the case of viscous spreading) this results in no balancing of the disc with the external photoevaporative wind, and so the disc continues to truncate over the disc lifetime.

Comparing fig. \ref{fig:surface_density} with figure 6 of \citet{Tabone22}, the wind dominated disc evolves in a similar manner with it constantly reducing in size and mass.
However, whilst the viscous case here truncated to an equilibrium, the equivalent case in \citet{Tabone22} continued to expand in size as it lost mass.
This shows the importance of including, even to a small degree (e.g. 10$\rm G_0$), the effects of external photoevaporation.

\begin{figure}
\centering
\includegraphics[scale=0.5]{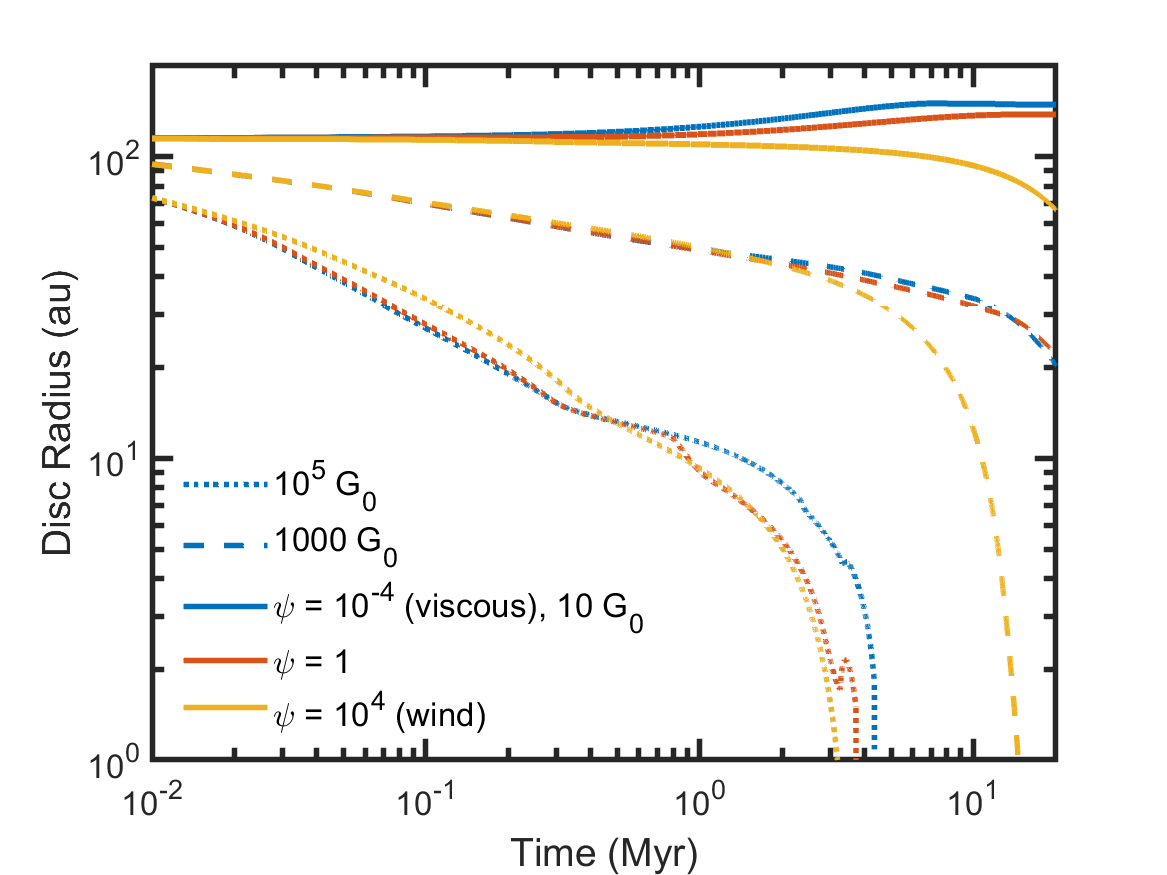}
\caption{Evolution of the disc radius taken at the radius that encompasses 90\% of the disc mass for discs evolving in different FUV environments: 10 $\rm G_0$ (solid lines), $10^3 \rm G_0$ (dashed lines) and $10^5 \rm G_0$ (dotted lines). The main disc evolution processes are shown by the colours: viscous discs (blue lines), hybrid discs (red lines) and wind driven discs (orange lines).}
\label{fig:r90}
\end{figure}

\subsection{Disc radii}

In terms of the disc radius, it is clear that there is a difference in evolution between viscous and wind dominated discs, with the wind dominated discs truncating at a much faster rate.
Figure \ref{fig:r90} shows the evolution of the disc radius, defined as that containing 90 per cent of the mass, for the viscous (blue line), hybrid ($\psi = 1$, red line) and wind (orange line) dominated discs. The solid, dashed and dotted lines are for external FUV radiation fields of 10, $10^3$ and $10^5$\,G$_0$ respectively. 
For low $\rm G_0$ environments, shown by the solid lines, it is clear that the wind dominated case has a different evolution track to those including the effects of viscosity, with it constantly decreasing in size.
The viscous and hybrid cases however expand outwards initially until they reach an equilibrium with the external mass loss rates, before they then slowly begin to truncate.
This is similar to that seen in other works \citep[e.g.][]{Tabone22} and should observations give hints to disc sizes over time, it could be possible to determine the driver of angular momentum transfer in protoplanetary discs.

However, in stronger environments, where mass loss rates are significantly higher, external photoevaporation is able to truncate the disc and dominate the discs evolution irrespective of the mode of angular momentum transport, typically from the outside-in \citep{Coleman22}.
The dashed and dotted lines in fig. \ref{fig:r90} show the evolution of disc radius for stronger FUV environments, that being 1000\,$\rm G_0$ and $10^5\,\rm G_0$.
With stronger mass loss rates, the equilibrium point between viscous expansion and mass lost through external photoevaporation moves inward, as can be seen by the steady decline in disc radii for the blue and red dashed lines in fig. \ref{fig:r90}, showing the evolution of discs in a 1000 $\rm G_0$ environment.
Interestingly there is no expansion in these discs, with the equilibrium point being located closer than that for the initial disc.
Equally of note, there is little difference between the viscous and wind dominated discs here for the first $\sim2$Myr, with the only main difference in the end being the faster truncation of the disc in the latter stages of the disc lifetime, when the discs are $<40\au$ in size.
This would indicate that maybe, only in older stellar clusters and with smaller compact discs, will it be possible to discern whether discs are driven through viscosity or through disc winds.

Increasing the external field strength further, with the dotted lines showing for a $10^5 \rm G_0$ environment, it is clear that all the discs in this case truncate down effectively to $\sim 10\au$ where the effectiveness of external photoevaporation strongly decreases, as the gas is too deep within the stars gravitational well.
The later evolution is then determined through either viscosity or disc winds, with those evolving through disc winds doing so at a faster rate, as the viscous case still attempts to maintain outward expansion, thus reducing the flow rate of material through the disc on to the star.
Evidently here, the differences in disc sizes is extremely small, making discs in strong FUV environments, poor targets to test other evolution processes of protoplanetary discs.

\begin{figure}
\centering
\includegraphics[scale=0.5]{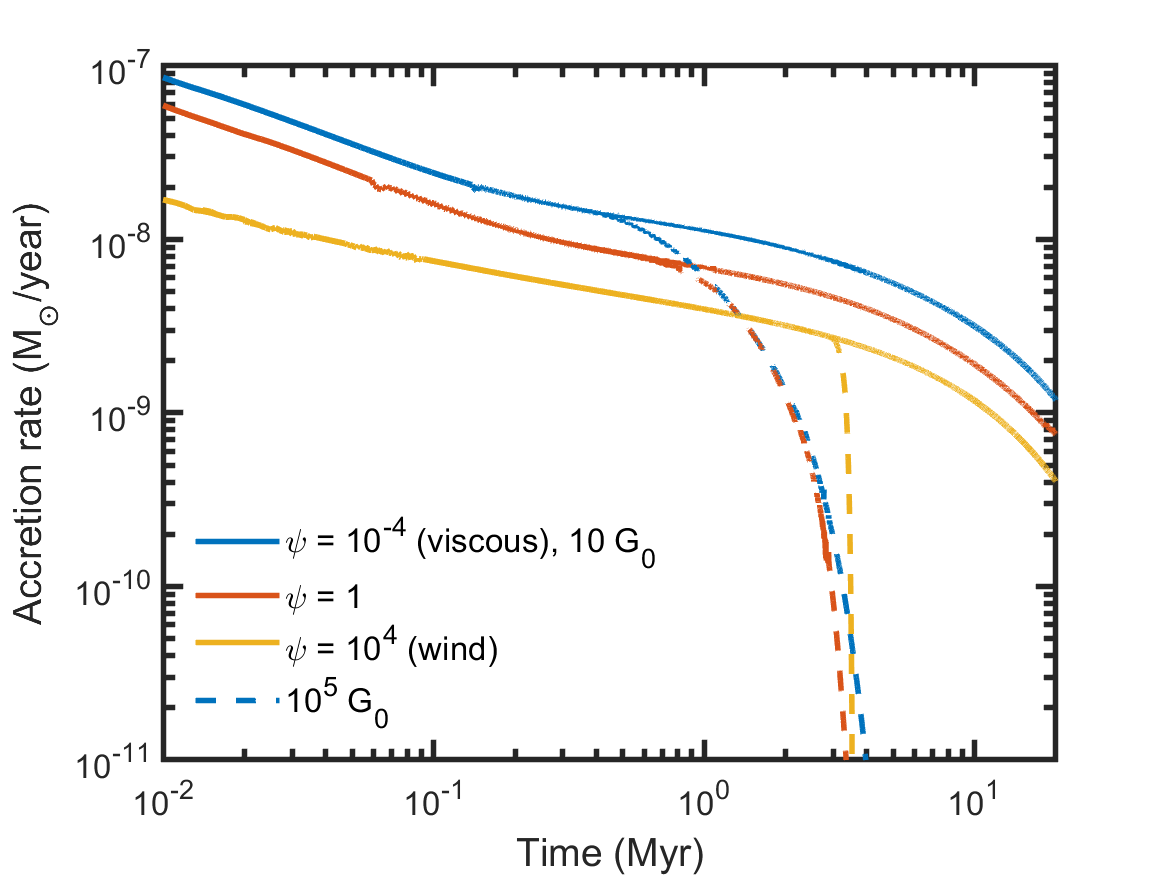}
\caption{Mass accretion rates on to the central stars for discs evolving in different FUV environments: 10 $\rm G_0$ (solid lines) and $10^5 \rm G_0$ (dashed lines). The main disc evolution processes are shown by the colours: viscous discs (blue lines), hybrid discs (red lines) and wind driven discs (orange lines).}
\label{fig:mdot_mix}
\end{figure}

\begin{figure*}
\centering
\includegraphics[scale=0.39]{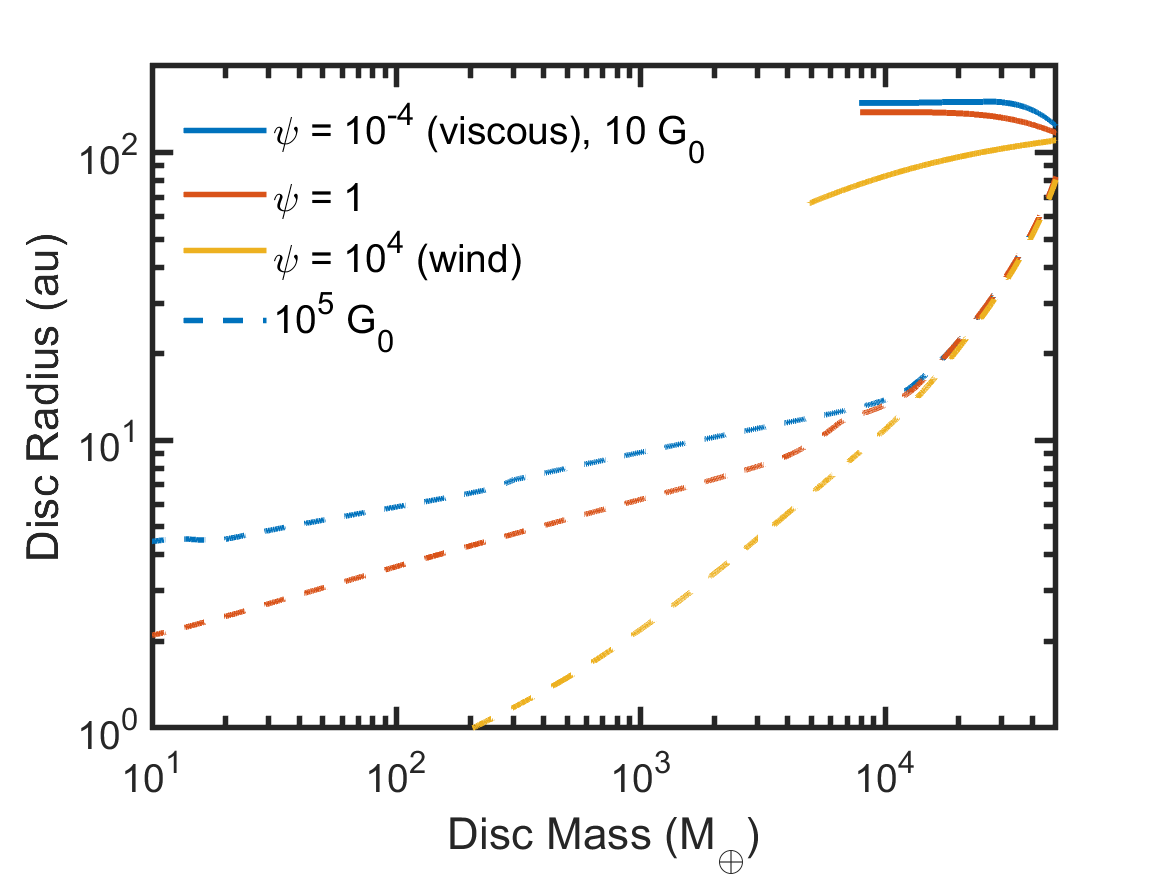}
\includegraphics[scale=0.39]{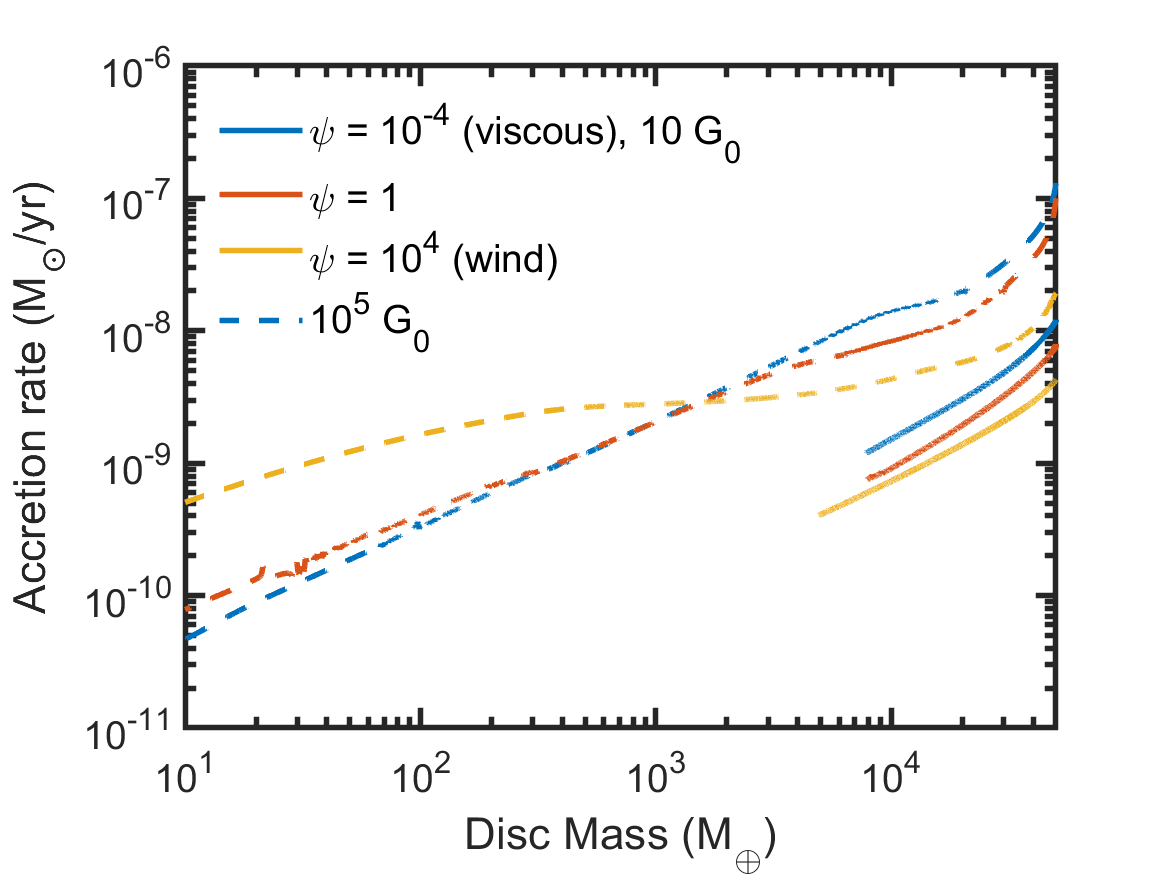}
\includegraphics[scale=0.39]{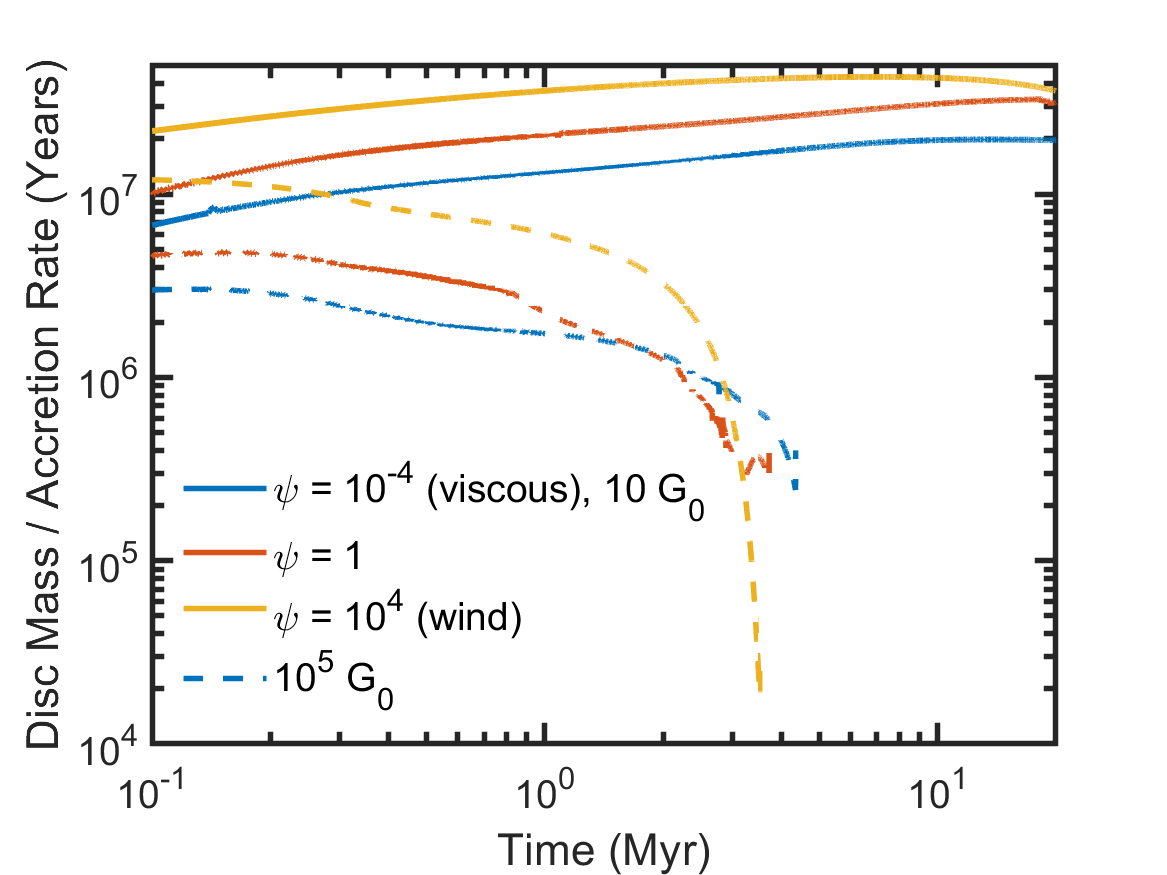}
\caption{Evolution of disc mass versus disc radii (left panel), disc mass versus accretion rate (middle panel), and the temporal evolution of the remaining disc lifetime (disc mass / accretion rate). Solid lines show discs evolving in $10 \rm G_0$ environments, whilst dashed lines show discs evolving in $10^5 \rm G_0$ environments. Viscous discs are shown with blue lines, whilst wind driven discs are denoted by orange lines, with hybrid discs shown by red lines.}
\label{fig:triple_mix}
\end{figure*}

\subsection{Mass Accretion Rates}

Whilst the disc radius shows some environments that could yield observable differences between viscous and wind driven discs, the mass accretion rate on to the star provides another empirical constraint on disc evolution \citep[e.g.][]{Manara23}.
In fig. \ref{fig:mdot_mix} we show the mass accretion rates over time for viscous (blue), hybrid (red), and wind driven (orange) discs in a weak environment of 10$\rm G_0$ (solid lines), and a strong environment of $10^5 \rm G_0$ (dashed lines).
The wind driven case can be seen to be around a factor few times smaller than the viscous discs for most of the disc lifetime. This is irrespective initially of the environment, since the material in the inner disc is not affected by the local environment until it begins to affect the resupply of material from the outer disc.
At this point, the mass accretion rates drop, to balance what mass is flowing through the disc.
Interestingly for the stronger environment, the more viscous discs evolve faster and reduce in accretion rate quicker than the wind driven discs.
This is mainly due to the balancing of viscous expansion, limiting the supply of gas to the inner regions of the disc, to be accreted on to the central star.
Since there is no expansion for wind driven discs, this allows them to continue accreting through the disc at a similar rate until, the end of the disc lifetime when there is no material remaining and the accretion rate plummets dramatically.

Whilst there are some subtle differences in the mass accretion rates, in practice this is likely to be a prohibitively difficult individual measure to differentiate between viscous and wind driven discs.
Since the accretion rate is dependent on the inner regions of the disc, it is therefore heavily dependent on the initial disc properties.
These include the disc mass, the disc size and how compact it is, as well as the strength of the MHD wind or turbulence. It may also be sensitive to the presence of planets within the inner disc and internal photoevaporation (as we discuss below). 
Manipulating the initial conditions well within reasonable bounds, would yield similar mass accretion rates, washing out any discernible difference between viscous and wind driven discs.
This is in contrast to the disc radius, where the qualitative evolution of the different processes would remain.

\subsection{Other Observables}

We now look at a number of other observable properties that are thought to give rise to differences between viscous and MHD wind evolving protoplanetary discs.
Figure \ref{fig:triple_mix} shows the relation between disc mass and disc radius (left panel), disc mass and accretion rate on to the central star (middle panel), and the temporal evolution of disc mass divided by the mass accretion rate, a proxy for the disc lifetime (right panel).
The colours again show a viscous disc (blue), a hybrid disc (red) and a wind driven disc (orange), whilst solid lines show the discs evolving in a weak environment, and dashed lines show a strong environment.
Expectations from analytical models show a divergence in how disc radii evolve with disc mass, with viscous discs becoming larger in radius as the discs evolve and lose mass \citep{Manara23}.
For wind driven discs, they continue to reduce in size as mass in accreted on to the central star.
Our models here are somewhat in agreement, with viscous models generally being larger than wind driven models for similar disc masses, but the expansion of the viscous discs here is balanced by (even weak) external photoevaporation, before they begin to slowly reduce in size.
Additionally, whilst those discs in stronger FUV environments also show a similar trend, the differences are only noticeable once the discs are less than 10$\au$ in size, and so such small differences will be difficult to realise in observations.

Moving to the middle panel of fig. \ref{fig:triple_mix}, analytic expectations for viscous discs have found correlations between the disc mass and the mass accretion rate, where discs tend to have lower accretion rates as the discs reduce in mass, and reach the end of their lifetime \citep{Jones12,Lodato17}.
When including the effects of internal photoevaporation, \citet{Somigliana20} found that instead of a linear correlation between the mass accretion rate and disc mass, there was instead a ``knee'' when mass accretion rate fell below the photoevaporative mass loss rate, drastically reducing the mass accretion rate. In analytic comparisons to wind driven discs, this results in the viscous discs having lower mass accretion rates than wind driven discs for discs of similar mass \citep{Manara23}.
Looking at fig. \ref{fig:triple_mix}, this is not seen in the weak external environment (solid lines) where the viscous disc maintains larger accretion rates than the wind driven disc. However these discs have not reached the end of their lifetime, since we stopped the simulations after 20 Myr, where the effects may be observed.
For the stronger external environment (dashed lines), such a trend where the accretion rate for viscous discs is lower is observed, but only once the discs are reduced to $\sim 1-2$ Jupiter masses, and by comparing to the left panel of fig. \ref{fig:triple_mix}, this is only once they are $\sim$few $\au$ in size.
However, no such knee feature is observed, since our models shown in fig. \ref{fig:triple_mix} did not include internal photoevaporation.
Whilst the differences are not large, only a factor few at most, these again could be matched by altering the initial properties of the disc.
However, given the compactness of the discs in the wind driven case, this would suggest that they will have significantly larger accretion rates at smaller disc sizes than those that evolve through viscous accretion.
Should observations be able to accurately obtain both measures for compact, weakly accreting discs, this could shed light on the processes that occur within them.
It is also interesting to note that in stronger external environments, both evolving viscous and wind driven discs exhibit larger accretion rates than those of similar mass in weak environments. This is expected since the accretion rate is more affected by the inner disc region, whilst the disc mass is influenced by the processes in the outer disc.
Therefore, observations of different star forming regions should observe this signature.

Finally, by using a proxy for the disc lifetime, though note it assumes no other mechanism for mass loss, the right panel of fig. \ref{fig:triple_mix}, shows the temporal evolution of the disc mass divided by the accretion rate, i.e. the depletion time-scale.
From previous works \citep[e.g.][]{Jones12,Somigliana23,Manara23}, it is expected the remaining lifetime of viscous discs increases as time progresses. This is mainly due to the assumptions that the discs are constantly expanding, whilst accretion rates are falling.
However even with the inclusion of weak external photoevaporation this expansion is halted, making the assumption no longer valid.
Indeed, even the solid lines for a weak external environment show the remaining lifetime remaining level, similar to the wind driven case.
In fact, the remaining lifetime, by this metric is also shorter for the viscous discs.
When increasing the strength of the local environment, this again does not agree with analytic expectations, as both the viscous and wind driven rates very quickly drop as the discs are mainly dispersed through photoevaporative winds.
Only when the inner disc remains, does the remaining lifetime in the viscous disc exceed that of the wind driven, as it tries to simultaneously expand whilst accreting on to the central star.
Therefore in summary, when observing discs in regions where even weak external photoevaporation is present it will be challenging to use this metric to differentiate between viscous and MHD wind driven discs.

\begin{figure}
\centering
\includegraphics[scale=0.5]{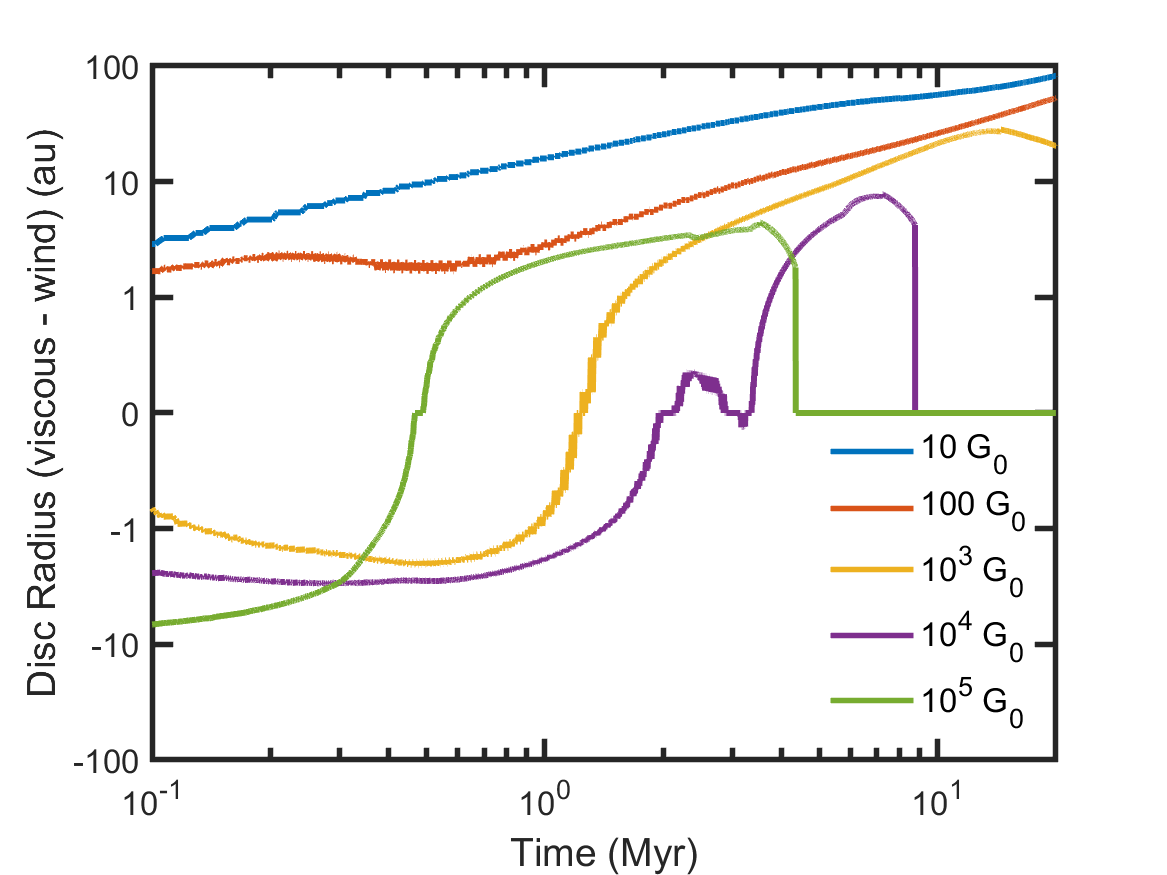}
\caption{Differences in disc radii over time between a viscously evolving disc and a wind driven disc. Different FUV environments are denoted by colours: $10\rm G_0$ (blue), $100\rm G_0$ (red), $1000\rm G_0$ (orange), $10^4\rm G_0$ (purple) and $10^5\rm G_0$ (green).}
\label{fig:radius_diff}
\end{figure}

\subsection{Optimal observations}

With the above sections showing how viscous and wind driven discs evolve in different environments, we now explore the regions in time and space that show the largest differences for such discs.
Figure. \ref{fig:radius_diff} shows the difference in disc radii over time between viscous discs and wind driven discs, for a range of external FUV radiation field strengths.
For values $>0$ in fig. \ref{fig:radius_diff}, viscous discs are larger than wind-driven discs and vice versa for values $<0$.
It is clear from fig. \ref{fig:radius_diff} that discs that evolve in stronger FUV environments ($\ge 10^4\,\rm G_0$) show very little difference in disc radii (i.e. $<10\au$).
Additionally in more intermediate environments, e.g. $10^3 \rm G_0$, viscous discs are only more than 10$\au$ larger, after $\sim 7$Myr, when the discs have evolved somewhat and will be smaller in size.
In weaker environments, viscous discs are nearly always larger, but these differences are only greater than 10$\au$ after 1--2 Myr.
This effectively shows the age of the disc that observations are required to yield the greatest differences between viscous and wind driven discs.

\begin{figure}
\centering
\includegraphics[scale=0.5]{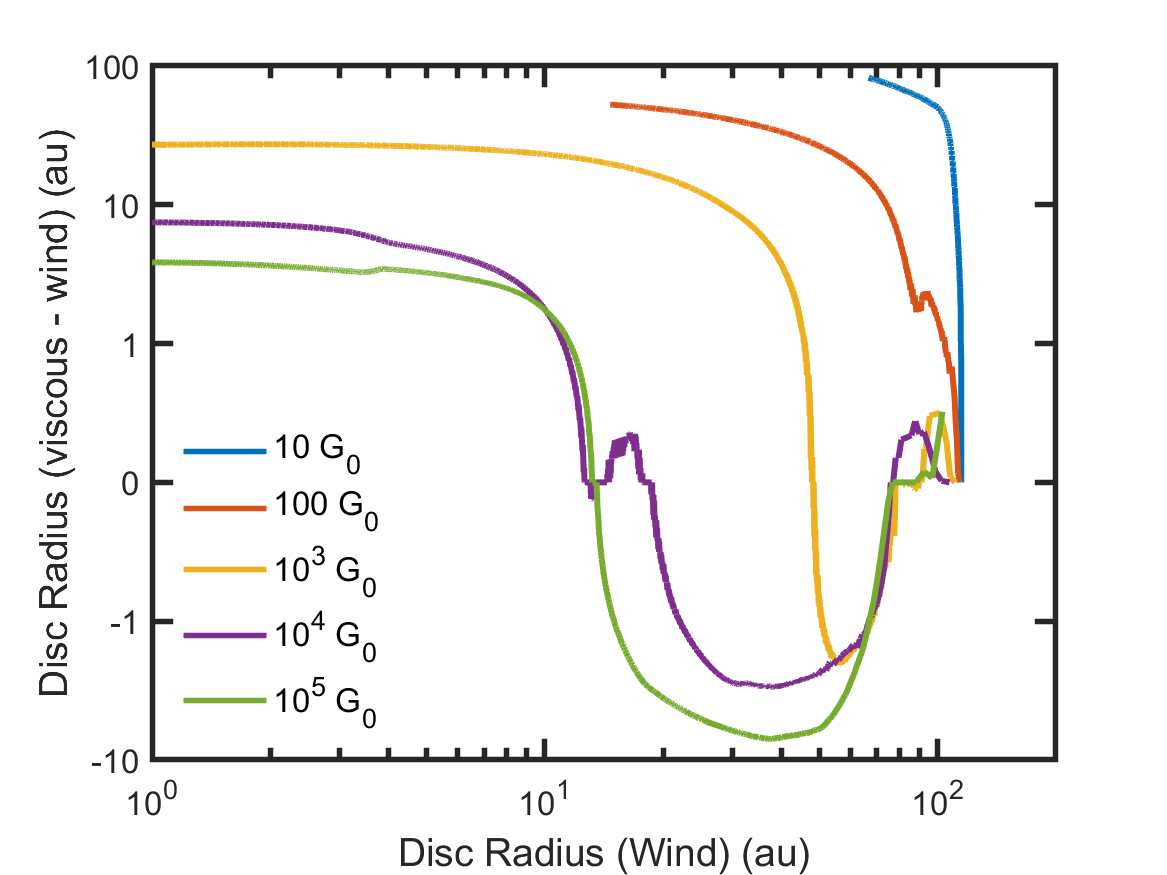}
\caption{Differences in disc radii as a function of the wind driven discs radii between a viscously evolving disc and a wind driven disc. Different FUV environments are denoted by colours: $10\rm G_0$ (blue), $100\rm G_0$ (red), $1000\rm G_0$ (orange), $10^4\rm G_0$ (purple) and $10^5\rm G_0$ (green).}
\label{fig:radius_diff_rad}
\end{figure}

Whilst fig. \ref{fig:radius_diff} showed the difference in disc radii over time, it is also interesting to know how this relates to the disc size.
Figure. \ref{fig:radius_diff_rad} now shows the disc radius against the difference in radii between viscous and wind driven discs.
It is clear that for weaker environments, shown by the blue and red lines, the discs are fairly extended when there is a noticeable difference.
However for the intermediate environment, the orange line showing a disc in a $10^3 \rm G_0$ environment, whilst the differences become $\ge10\au$ after $\sim 7$Myr, this is only when the wind driven disc is 30$\au$ in size.
This means that knowing the disc age, and other properties that influence its evolution will become important in comparing viscous and wind driven disc models.
Additionally for the discs in strong UV environments, the main differences are only when the discs are extremely compact, and so where external photoevaporation is ineffective, and so this is dependent on how quickly the discs evolve to this state and the viscous discs become flat and slow accretors.
In summary, fig. \ref{fig:radius_diff_rad} again shows that weak environments would be the best targets for comparing disc sizes to determine whether discs are viscous or wind driven.

\begin{figure*}
\centering
\includegraphics[scale=0.5]{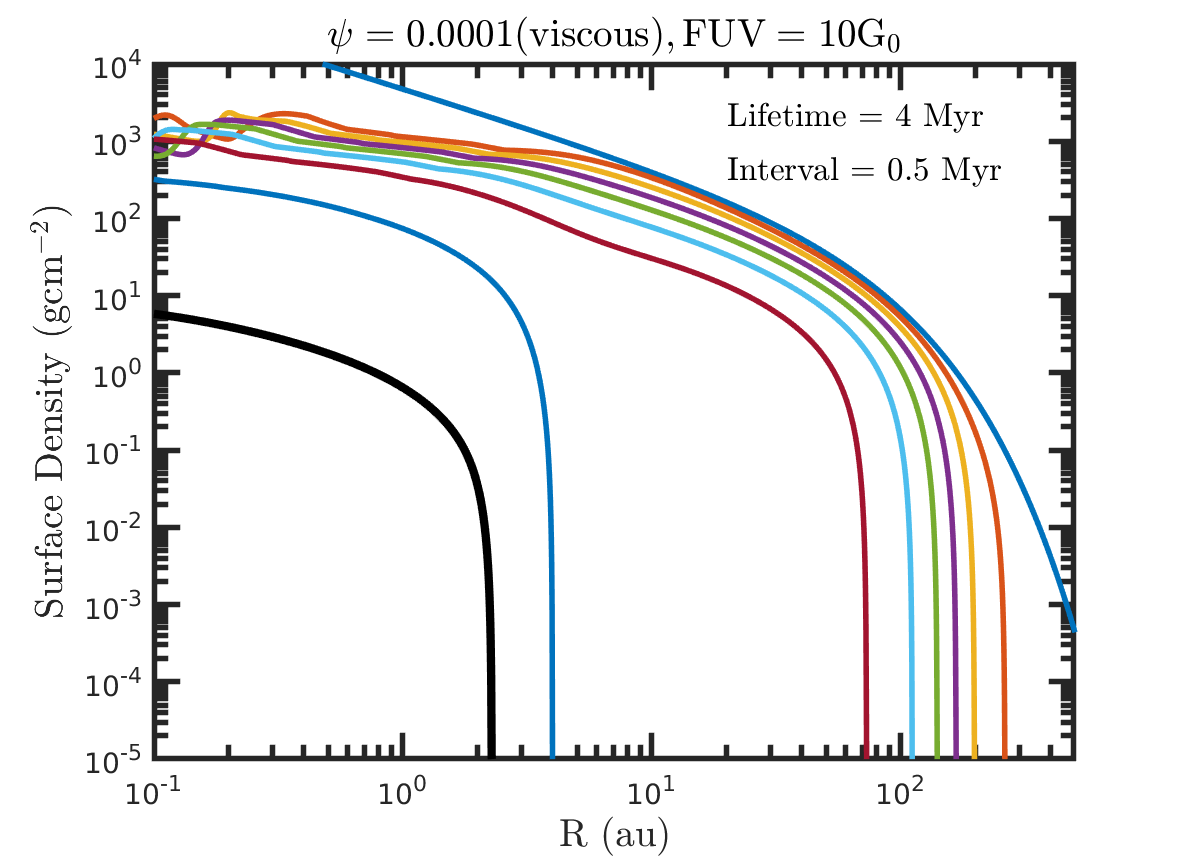}
\includegraphics[scale=0.5]{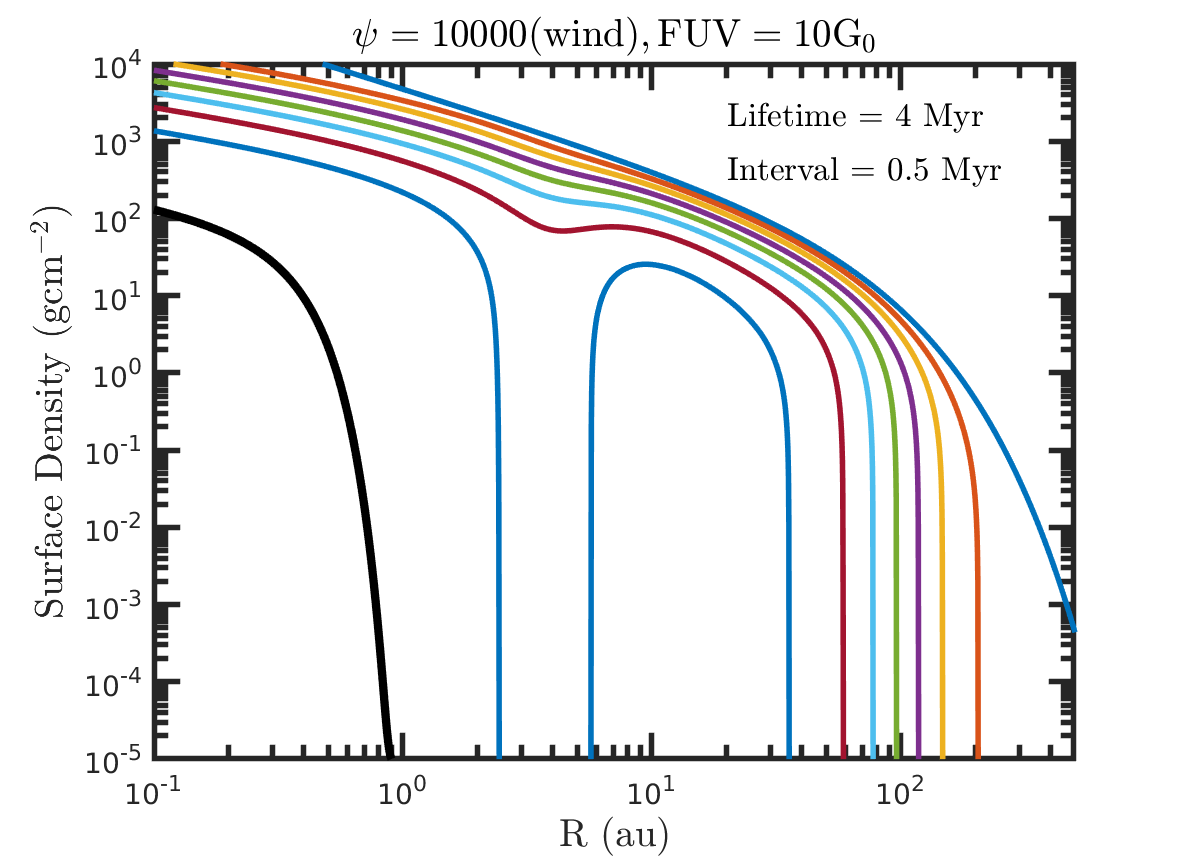}
\caption{Temporal evolution of gas surface densities for viscous discs (left panel) and wind-dominated discs (right panel), whilst including internal photoevaporation with the central stars X-ray luminosity $L_{\rm X} = 10^{30} \rm erg s^{-1}$.}
\label{fig:surface_density_IPE}
\end{figure*}

\section{Consequences of Internal Photoevaporation}
\label{sec:internal_pe}

Section \ref{sec:results} has shown how external photoevaporation affects viscous and wind driven discs, and indicated that discs in weak FUV environments are best suited for observations with the aim of differentiating between viscous and wind driven disc evolution.
However, in these environments, the models show that the disc lifetimes are extremely long ($>10$\,Myr) much longer than those observed in nearby clusters \citep[e.g.][]{Haisch01, Ribas15}.
Additionally, the value of $\alpha$, used to determine the level of turbulence in viscous discs, was set to $10^{-3}$, which is consistent, if not slightly higher than that observed in protoplanetary discs \citep{Ansdell18,Villenave22}.
This value was also used for the strength of the MHD wind.
To obtain shorter lifetimes, increasing the strength of $\alpha$ would reduce the lifetime, however, this would be in disagreement with observational constraints on $\alpha$.
Also, for wind driven discs, this would increase the mass accretion rate, which again would need to be compared with observed values.
Other ways of reducing the lifetime would be on altering the initial properties, of the disc, i.e. starting with smaller, less massive discs.
However, this would not solve the viscous disc evolution sufficiently, since viscous expansion would still lead to an equilibrium point with external photoevaporation, that would leave large abundances of mass in the outer disc, which would be slowly removed from the system. Additionally, whilst wind driven discs will be sufficiently more compact as well, and with no process for expansion, they may be too compact to be accurately compared with observations.

We have so far not considered internal photoevaporation in this paper.
Energetic photons from the central star, typically X-rays, can act to heat the surface layers of the disc and drive a photoevaporative wind.
For Solar mass stars, this wind typically removes mass from 10-100 $\au$, a similar region to the MHD wind closer to the star, and the external photoevaporative wind far from the star.
The interplay between internal photoevaporative and MHD winds is not extensively studied \citep{2017ApJ...847...11W}, though recent work has studied the interplay between Hall-effect MHD and internal photoevaporation \citep{Sarafidou23}, finding that winds are mainly driven by photoevaporation except when the magnetic field is strong or the X-ray luminosity is low ($\log_{10}(L_X)\leq29.3$). Given the uncertainty in how to treat internal photoevaporation and MHD winds simultaneously, we only include a weak internal photoevaporative wind. Specifically we set the central star's X-ray luminosity $L_{\rm X}=10^{30} \rm erg s^{-1}$, whereas the average value for Solar mass stars in star forming regions is $L_{\rm X}=10^{30.5} \rm erg s^{-1}$ \citep{Flaischlen21}. Given we consider relatively weak X-ray luminosities, and given the uncertainty in the interaction between photoevaporative and MHD winds, we treat both the MHD and photoevaporative winds entirely separably, to make an initial exploration of their effects on the models.

\subsection{Surface Density Evolution}

Figure \ref{fig:surface_density_IPE} shows the gas surface density evolution for a viscous disc (left panel) and a wind driven disc (right panel).
Like fig. \ref{fig:surface_density}, these discs evolved in a weak FUV environment (10$\rm G_0$), but this time include internal photoevaporation due to X-rays.
Comparing to fig. \ref{fig:surface_density}, it is clear that the discs evolve substantially faster when there is even a weak internal photoevaporative wind.
Indeed the discs are fully dispersed after only 4 Myr, much closer to the typical  observed protoplanetary disc lifetimes of $\sim3$\,Myr \citep[e.g.][]{Haisch01, Ribas15}.
This is mainly due to the internal photoevaporative wind removing gas from the intermediate regions of the discs, which removes the supply of material to the outer disc, allowing external photoevaporation to truncate the disc at a faster rate.
Interestingly, both the viscous and wind driven discs had similar lifetimes, and it appears from fig. \ref{fig:surface_density_IPE}, similar evolution profiles.
Nearer to the end of the disc lifetime, internal photoevaporation would be able to open a hole in the disc, as seen in the right panel of fig. \ref{fig:surface_density_IPE}, where either viscous transport or the MHD wind would be unable to replenish the material at that radius lost through the wind.
Once this occurs, the disc radii can truncate extremely quickly, since as with internal photoevaporation truncating outwards on the inner edge of the outer disc, and external photoevaporation acting on the outer edge, eventually such an outer disc will become thin, ring-like, before being dispersed.
This would cause a sudden drop in the radius of the disc.
Note that the gap opened by internal photoevaporation in the viscous case is not shown in fig. \ref{fig:surface_density_IPE}, since the gap opened and additionally the outer disc quickly dispersed in between our plotting intervals. In both cases, the gaps in the discs were only open for a short time ($<0.1$ Myr), since the outer discs were quickly dispersed.
Interestingly, the gap was able to open at a slightly earlier time in the viscous case, since viscous expansion of the outer disc allowed for a larger external photoevaporative mass loss rate, that depleted the outer disc quicker. Whilst this did not affect the disc lifetime, since the final lifetime was determined by the rate of accretion in the inner disc, this did result in the outer disc dispersing at an earlier time.

\begin{figure}
\centering
\includegraphics[scale=0.5]{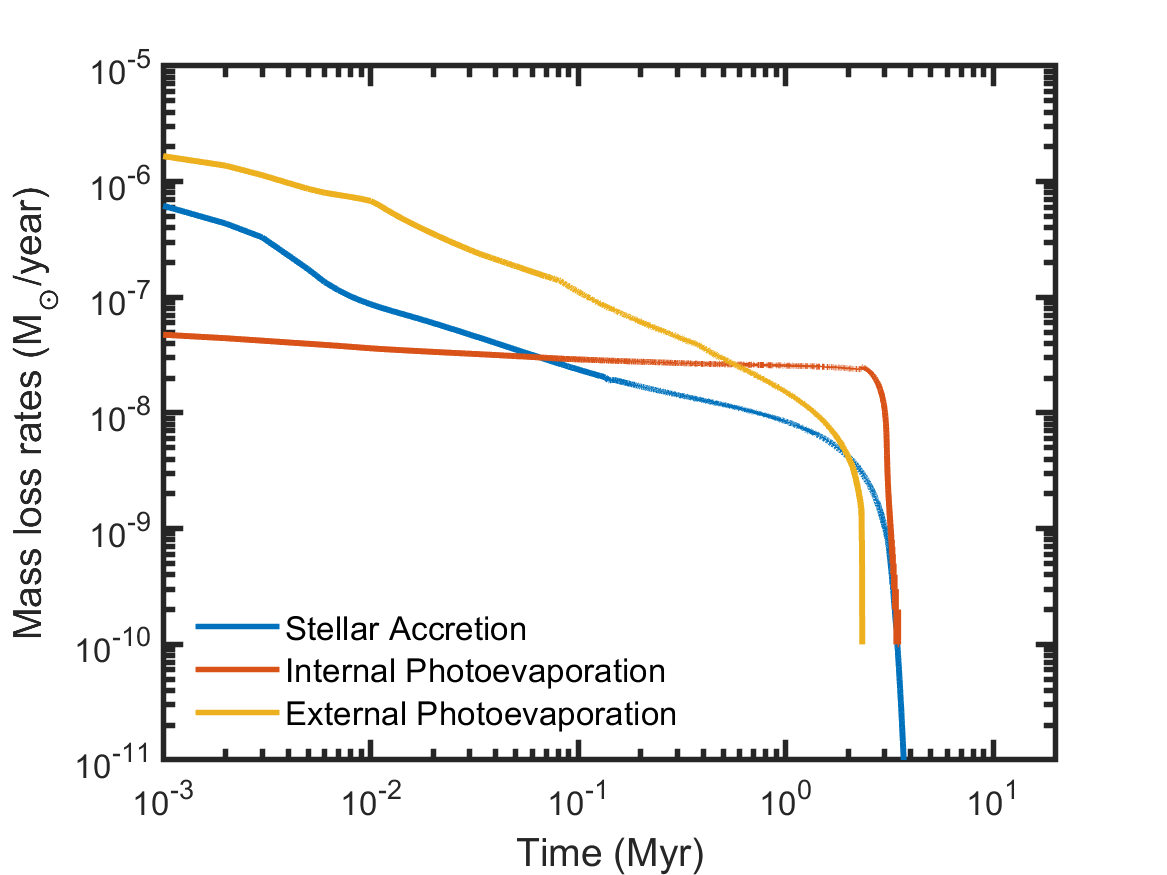}
\caption{Mass loss rates for viscous discs evolving in a $10^{3} \rm G_0$ environment, with contributions from stellar accretion (blue line), internal photoevaporation (red line) and external photoevaporation (yellow line).}
\label{fig:mdot_IPE}
\end{figure}

To further highlight the influence of photoevaporation on the evolution of the disc mass, fig. \ref{fig:mdot_IPE} shows the mass loss rates for a viscous disc in a $10^3 \rm G_0$ environment. The blue line shows the accretion rate on to the central star, whilst the red and yellow lines denote the mass lost through internal and external photoevaporation respectively.
The effect of external photoevaporation is clear in this environment, since it dominates the total mass loss rate of the disc for the first $\sim0.5$ Myr, whilst the disc is large. After this time, internal photoevaporation begins to dominate since it removes mass from the intermediate regions of the disc, where external photoevaporation is less effective. Interestingly, it is only early in the disc lifetime, that the stellar accretion rate is larger than the internal photoevaporation rate, since there is a large buildup of material in the inner disc. As the disc depletes, the accretion rate on to the star also decreases, consistent with observed trends in regions such as Chamaeleon I \citep{Manara16,Manara17} and Lupus \citep{Alcala14,Alcala17}.
After only $\sim0.1$ Myr, the mass accretion rate is weaker than the internal photoevaporative mass loss rate. Note that this is also consistent for discs in other external UV environments, and additionally from discs evolving through MHD disc winds. It is also worth noting that the central star's X-ray luminosity is weak here compared to the average observed in stellar clusters \citep{Flaischlen21}, and so we consider this a weak internal photoevaporative wind, as larger X-ray luminosities will yield much stronger mass loss rates.
Therefore, fig. \ref{fig:mdot_IPE} emphasises how photoevaporation dominates the mass loss mechanisms for protoplanetary discs, with the relative strengths between internal and external photoevaporation determining which mechanism dominates the overall mass loss contributions. This was also seen in recent work that included high photoevaporation rates \citep{Coleman22}, and further illustrates the importance of including photoevaporation within models of protoplanetary disc evolution.

\subsection{Effects on disc radii}
\begin{figure}
\centering
\includegraphics[scale=0.5]{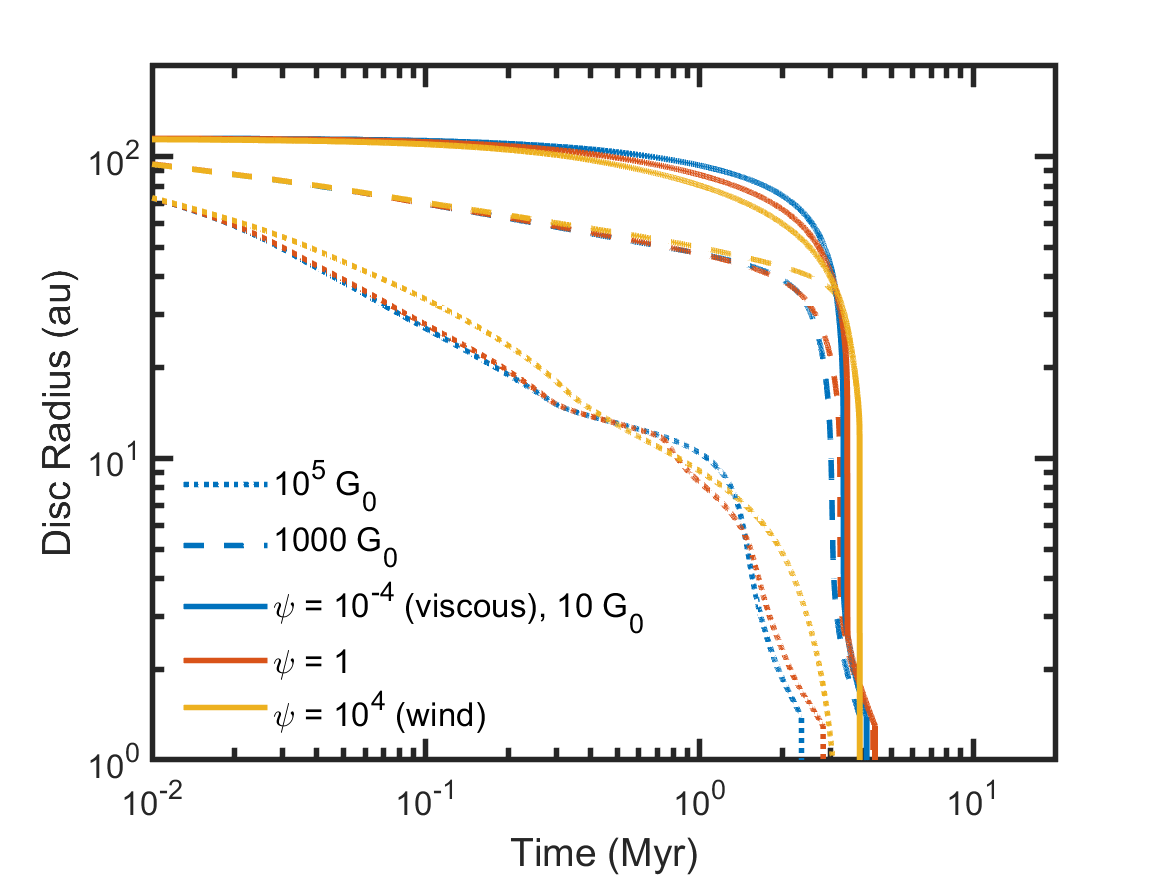}
\caption{Same as fig. \ref{fig:r90} but for discs including internal photoevaporation with the central stars X-ray luminosity $L_{\rm X}=10^{30} \rm erg s^{-1}$.}
\label{fig:r90_IPE}
\end{figure}

The similarities in disc radii is clearly evident in fig. \ref{fig:r90_IPE}, which shows the evolution of the disc radius, taken at 90 per cent of the disc mass, for viscous or wind driven discs in different environments.
Where in fig. \ref{fig:r90} there were differences between viscous and wind driven discs in the weak external environments, there are minimal differences here in fig. \ref{fig:r90_IPE} when even just a weak internal photoevaporative wind is included.
Additionally, discs are seen here to dramatically reduce in radius, this being the time just after internal photoevaporation has opened a hole in the disc, with the outer disc then being quickly dispersed from both sides.
Interestingly it may actually be this feature that is a useful indicator of whether discs are viscous or wind driven, since it appears that viscous discs evolve faster than wind driven discs in more intermediate environments.
Here the viscous expansion fed a stronger external mass loss rate, which ultimately allowed the outer disc region to deplete at a faster rate, once internal photoevaporation was able to open a hole.
This however is not the case for stronger external environments, since external photoevaporation there truncates the disc down to a small size before a hole is able to open \citep{Coleman22}.

In consequence, the inclusion of internal photoevaporation here has significantly dampened the possibility of using the disc radius as a measure of whether discs are viscous or wind driven.
This, along with the inclusion of external photoevaporation above, shows the importance when exploring what drives the evolution of protoplanetary discs, of including all of the major processes that determine the outcome.

\begin{figure*}
\centering
\includegraphics[scale=0.39]{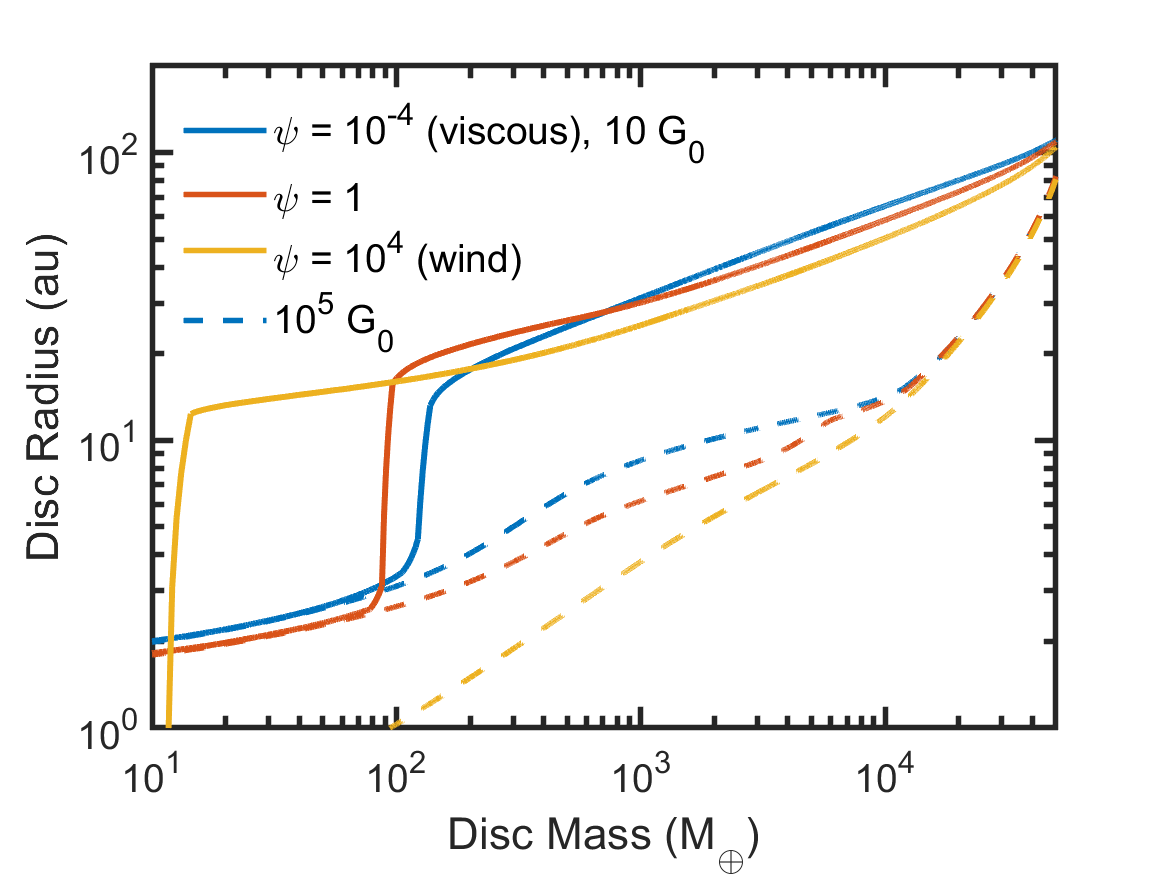}
\includegraphics[scale=0.39]{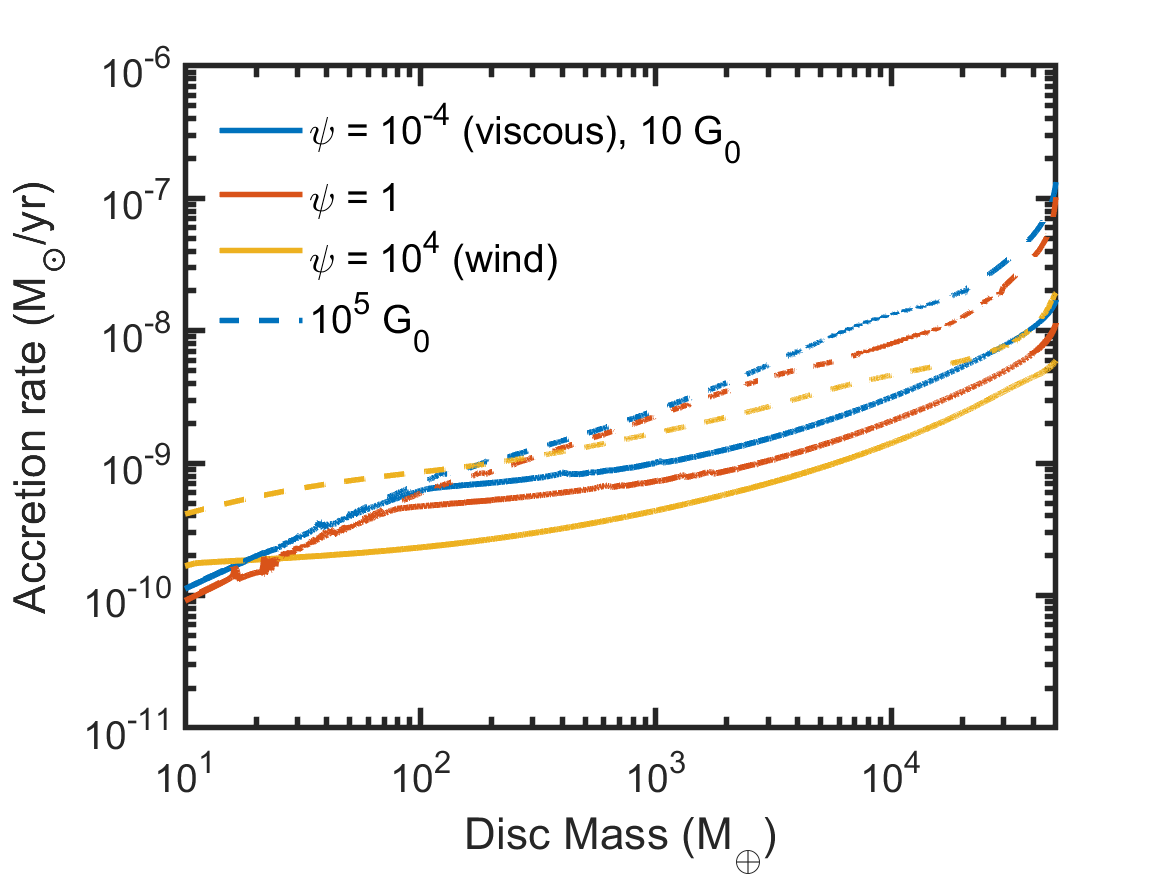}
\includegraphics[scale=0.39]{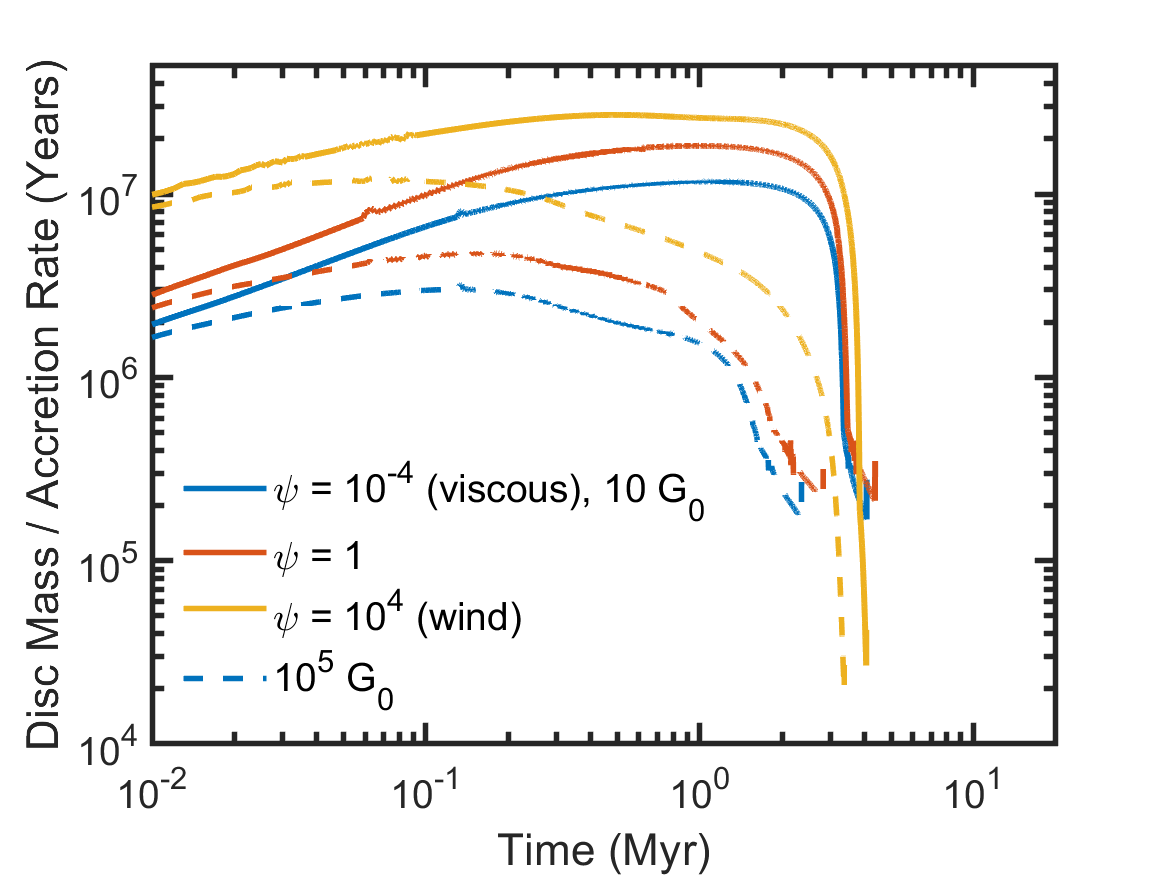}
\caption{Same as fig. \ref{fig:triple_mix} but for discs including internal photoevaporation with the central stars X-ray luminosity $L_{\rm X}=10^{30} \rm erg s^{-1}$.}
\label{fig:triple_mix_IPE}
\end{figure*}

\subsection{Other Observables}

Whilst the inclusion of internal photoevaporation has limited the effectiveness of using the disc radius as a measure between viscous or wind driven discs, it is also a question of how other observables, such as the disc mass, or the mass accretion rate, are also affected.
Figure. \ref{fig:triple_mix_IPE} shows the evolution between disc mass and disc radius (left panel), disc mass and accretion rate (middle panel), and time and remaining lifetime (right panel).
Similar to fig. \ref{fig:triple_mix}, the colours show viscous (blue), hybrid (red), and wind driven (orange) discs, with solid lines showing for a weak external environment, and dashed lines for a strong environment.

Looking at the left panel of fig. \ref{fig:triple_mix_IPE}, it is again clear that viscous discs are found to be slightly more massive than wind driven discs of similar size.
This is similar to what is observed in fig. \ref{fig:triple_mix}, however, the extent of the differences is now down to only $\sim$few $\au$, and so it is unlikely that such a difference could be unambiguously demonstrated observationally, due for example to uncertainty in the initial conditions of the disc.
For the discs in the weak external environments, internal photoevaporation dominated the mass loss rates of the disc, especially in the intermediate and outer disc regions. With this occurring for both viscous and wind driven discs, this is what led to the evolution tracks being similar. However, for the stronger external environments, external photoevaporation dominated the mass loss rates of the disc, effectively truncating the discs, and again leading to similar evolution tracks, until the discs became compact where the effectiveness of external photoevaporation was diminished.

A similar situation is found when looking at the mass accretion rate as a function of disc mass. Again the differences between viscous and wind driven discs is minimal, whilst there is limited evidence for a ``knee'' feature as predicted by internal photoevaporation models \citep{Somigliana20}, mainly due to the effects of photoevaporation on the outer disc regions.
Additionally, when including internal photoevaporation, the mass accretion rate is also similar in all environments, with a slightly increasing trend towards higher accretion rates in stronger external environments for discs of similar mass. The similarity across environments is due to the length of time taken for the effects of the outer disc to influence the inner disc regions, through the resupply of material that is lost to stellar accretion. For discs in stronger environments, they exhibit larger accretion rates since the outer disc is truncated more efficiently and so the mass of the disc reduces at a faster rate. The inner disc however, is minimally directly affected by photoevaporation and so has similar accretion rates on to the star at similar times early in its lifetime.

The biggest difference between fig. \ref{fig:triple_mix} and fig. \ref{fig:triple_mix_IPE}, is the right hand panel, where the remaining lifetimes now all exhibit similar pathways.
For both weak and strong external environments, the remaining lifetime in a wind driven disc is consistently larger than it's viscous counterpart.
Additionally the expectations in \citet{Somigliana23,Manara23} that the remaining lifetime in viscous discs increases over time is again not seen, highlighting the importance of including the additional processes that remove significant amounts of mass from the disc, i.e. photoevaporation.
At the very least it shows the importance of understanding the interplay between MHD disc winds and photoevaporative processes, which is only just starting to be explored \citep{Sarafidou23}.
The effect of external photoevaporation seen here by the stronger environments (dashed lines) shows smaller depletion time-scales than in weaker environments. This is again due to the outer disc being efficiently truncated and reduced in mass, whilst the inner disc continues to accrete on to the star at similar rates as discs in weaker environments. This leads to smaller depletion time-scales for discs in stronger environments. The main effect of internal photoevaporation here is by dominating the evolution of discs in weaker environments, both viscous and wind driven, and so causing them to have similar evolution tracks.

\begin{figure}
\centering
\includegraphics[scale=0.5]{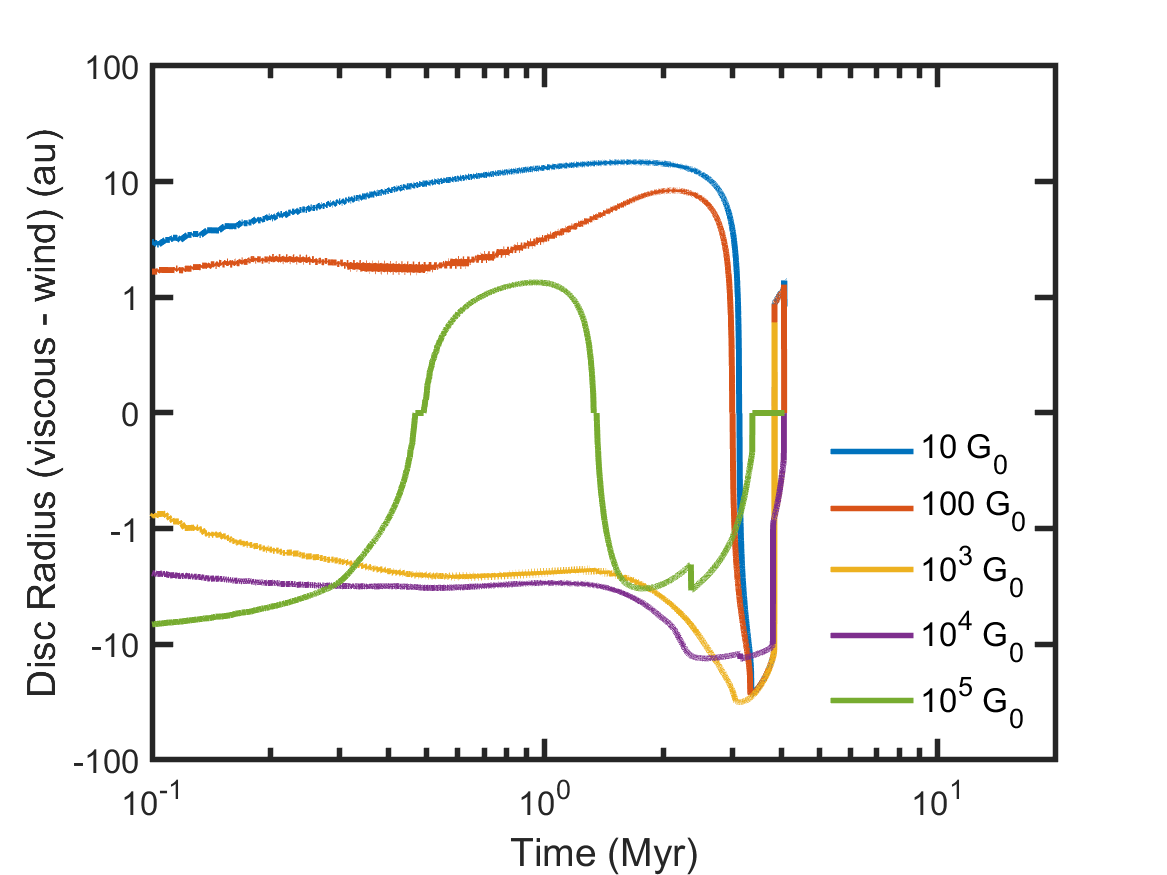}
\caption{Same as fig. \ref{fig:radius_diff} but for discs including internal photoevaporation with the central stars X-ray luminosity $L_{\rm X}=10^{30} \rm erg s^{-1}$.}
\label{fig:radius_diff_IPE}
\end{figure}

As was shown in fig. \ref{fig:r90_IPE}, there now appears to be minimal difference in the evolution of disc radii between viscous and wind driven discs.
Figure. \ref{fig:radius_diff_IPE} shows the difference in radius over time between viscous and wind driven discs for those evolving in different FUV environments (shown by the colours) and including the effects of internal photoevaporation.
Interestingly it seems that for the first Myr, there are now minimal difference across all environments.
Even in weak environments over the entire disc lifetime, the maximum difference is now 15 $\au$, showing that the discs are of similar size, irrespective of if they are viscous or wind driven, when also taking possible variations in initial conditions into account.
Interestingly the largest feature now is at the end of the disc lifetime, where wind driven discs in more intermediate environments (e.g. $10^3 \rm G_0$) are found to be significantly larger than their viscous counterparts.
This, as described above, is due to internal photoevaporation opening a hole in the disc, and coupled to the external wind, quickly disperses the outer disc.
With viscous expansion already supplying the external photoevaporative wind, there is less material in the outer disc to disperse than in the wind driven disc.
As such, this allows the outer disc to be more quickly removed, truncating the disc at a much faster rate.
This equally happens in stronger FUV environments, but here the discs fully truncate down to a small size before a hole is able to open, slightly nullifying the observable effect.

\section{Discussion}
\label{sec:discussion}

The above sections described the effects of external and internal photoevaporation on the evolution of viscous and wind driven protoplanetary discs, including the observables that can be measured from them.
We now look at other properties of the discs, to determine any other differences between viscous and wind driven discs under different processes.

\begin{figure}
\centering
\includegraphics[scale=0.5]{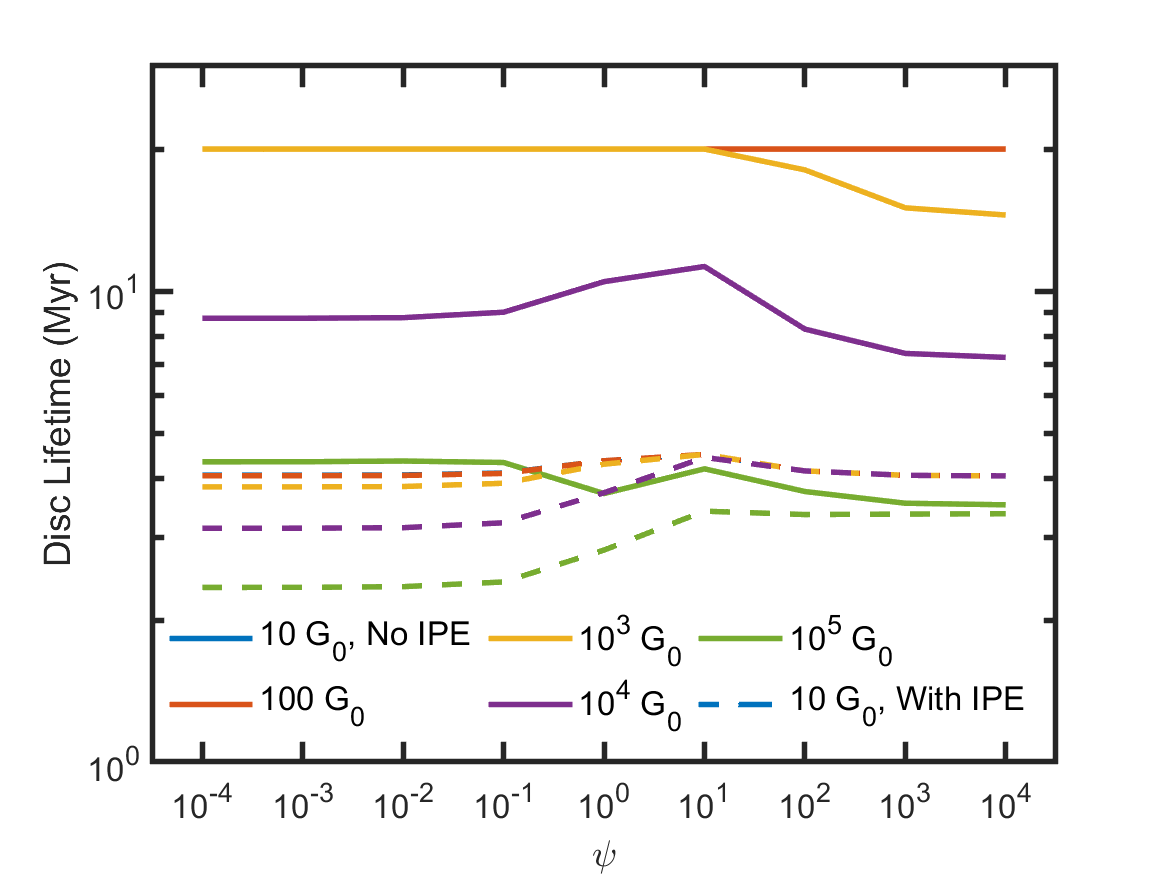}
\caption{Disc lifetimes as a function of $\psi$, the ratio of the strengths of viscosity and the MHD wind. Different environments are shown by colours: $10\rm G_0$ (blue), $100\rm G_0$ (red), $1000\rm G_0$ (orange), $10^4\rm G_0$ (purple) and $10^5\rm G_0$ (green). Discs not including internal photoevaporation are shown with solid lines, and those with internal photoevaporation are denoted by dashed lines.}
\label{fig:psi_lifetime}
\end{figure}

\subsection{Disc lifetimes}
One key observable is the disc lifetime, typically arising in observations from counting the fraction of stars in a cluster that exhibit signatures of a protoplanetary disc, e.g. through accretion on to the central star \citep[e.g.][]{Haisch01,Ribas15}.
Given that the discs have been shown to have some differences in their evolution profiles, we explore here the lifetimes that discs contain when either viscous or wind dominated.
Figure \ref{fig:psi_lifetime} shows the lifetimes of discs as a function of $\psi$. The strength of the external environment is denoted by the colour, whilst solid lines show the lifetimes for discs without internal photoevaporation, and dashed lines include internal photoevaporation.
The effect of internal photoevaporation on the disc lifetimes can be easily seen here, with the maximum lifetime now being roughly equal to $\sim$4 Myr when it is included.

In comparing the left side (viscous dominated) of fig. \ref{fig:psi_lifetime} to the right side (wind driven) it is clear that when internal photoevaporation is not included, disc lifetimes are shorter for wind driven discs. This is due to the viscous expansion slowing the accretion on to the star, as seen in the evolution of mass and radii in the figures above.
On the other hand, when internal photoevaporation is included, this trend flips, with viscous discs now having shorter lifetimes, as the expansion feeds both the internal and external photoevaporative winds.
This viscous expansion transfers material from the intermediate regions of the disc, i.e. 10--40$\au$, towards the outer regions of the disc, which allows for external photoevaporation mass loss rates to be stronger. The stronger external photoevaporation rates then deplete the outer disc quicker, and with the supply equally diminishing over time, as a result of the expansion and internal photoevaporation removing the material from this region, this led to the viscous discs losing mass more quickly, and thus having shorter lifetimes. With no such outward movement of material in wind driven discs, the mass loss rates due to external photoevaporation would be weaker than their viscous counterparts of similar ages, leaving internal photoevaporation to remove the majority of this material. This then results in the discs being more massive for longer, and thus having slightly longer disc lifetimes. This can be seen in the differences in the lifetimes in fig. \ref{fig:psi_lifetime}, by comparing the disc radii in fig. \ref{fig:r90_IPE}, as well as by comparing similar colour profiles in fig. \ref{fig:surface_density_IPE}.
Given that the opposite effects in terms of which mechanism yields longer disc lifetimes are seen here, this again highlights the importance of understanding the interplay between MHD disc winds and internal photoevaporation, and in particular, if and how they can work in tandem.

\begin{figure*}
\centering
\includegraphics[scale=0.39]{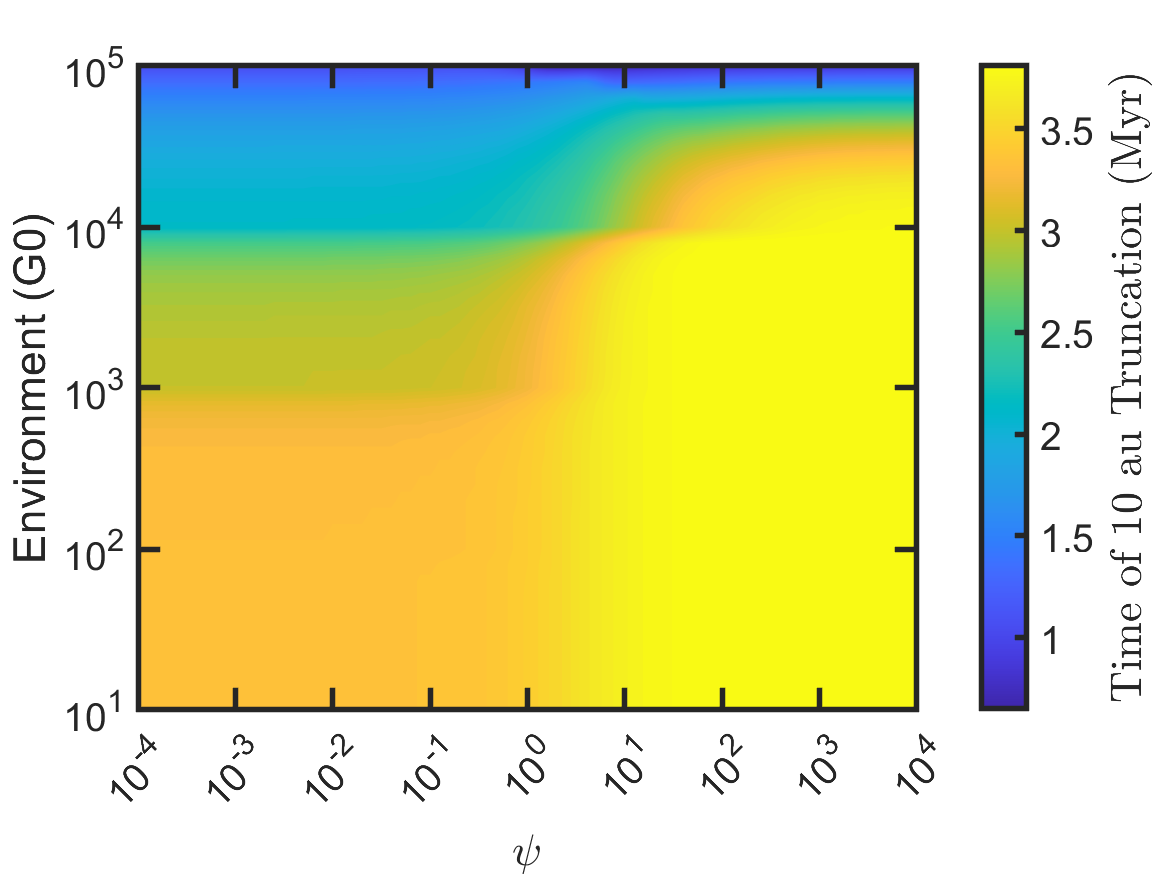}
\includegraphics[scale=0.39]{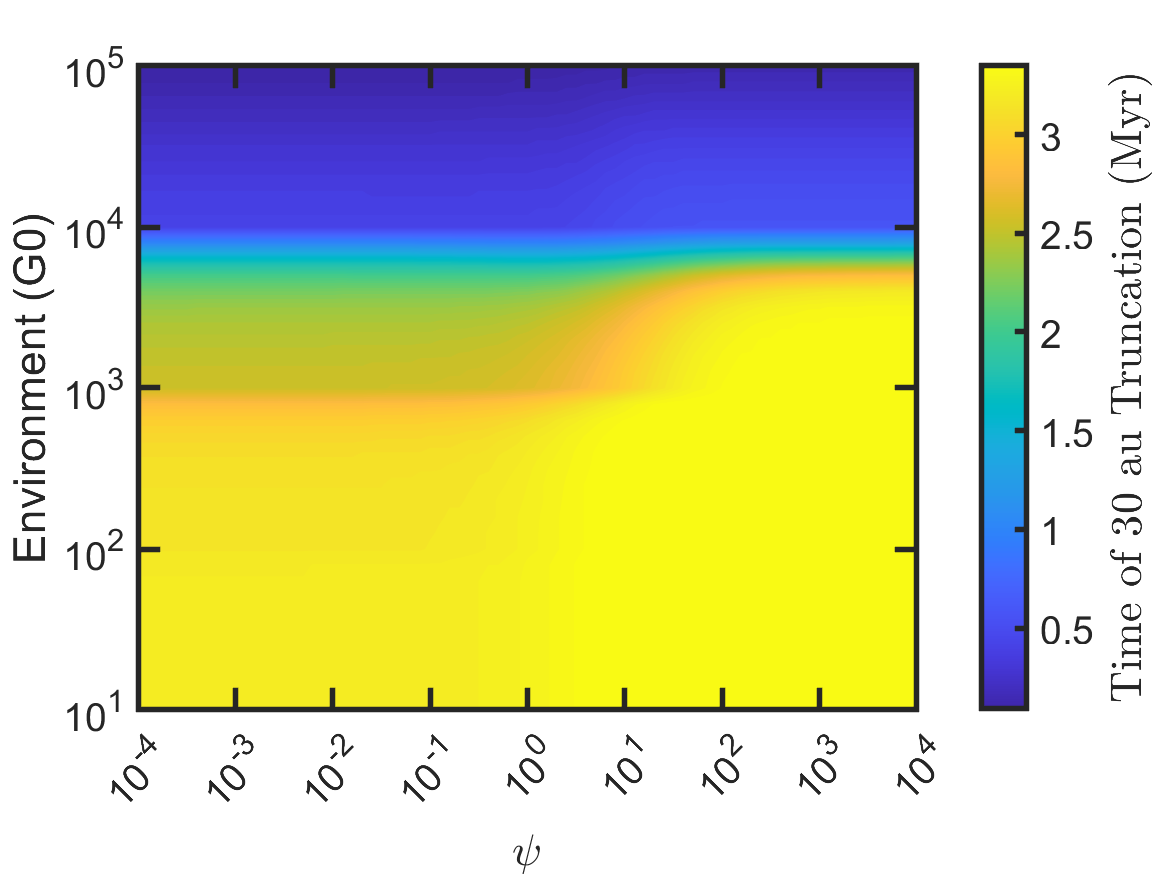}
\includegraphics[scale=0.39]{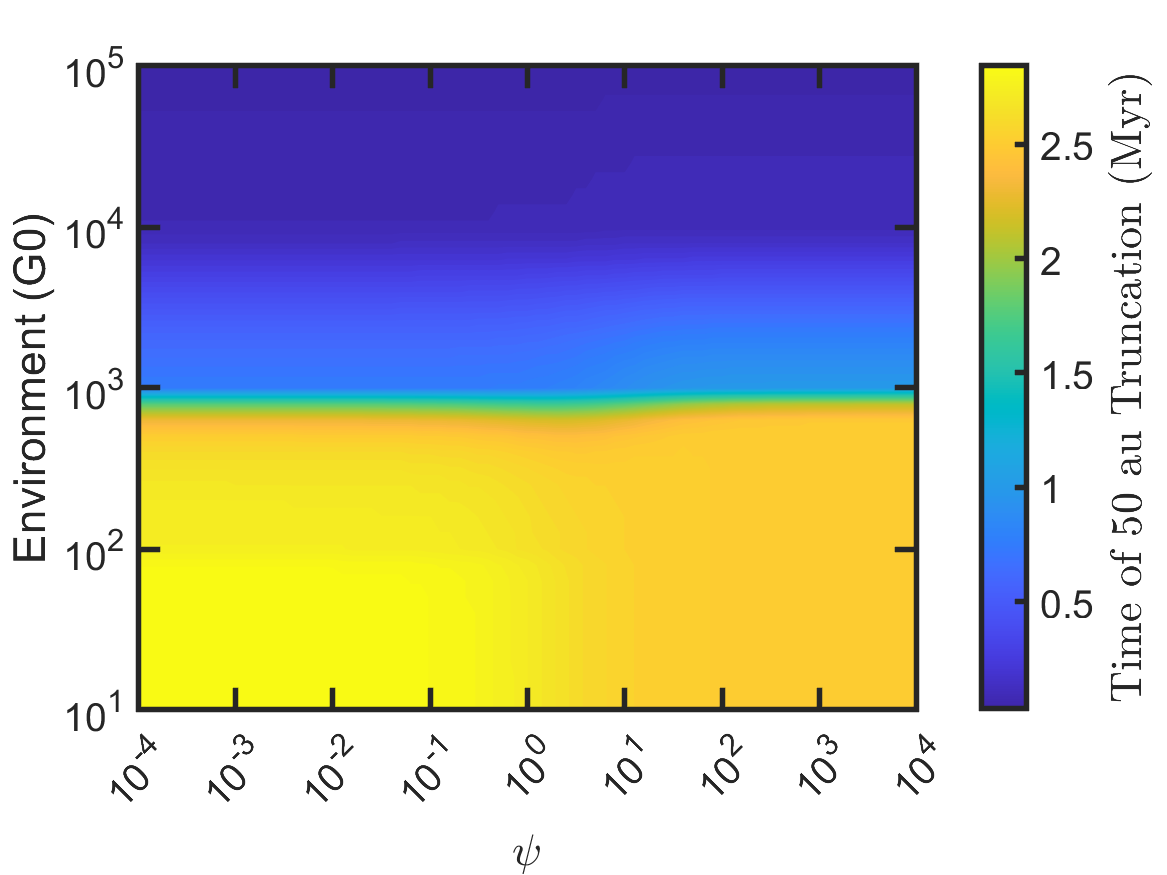}
\caption{Contour plots showing the truncation times to specific disc radii for discs in different environments (y-axis) and with different strengths of viscosity and MHD winds, $\psi$ (x-axis). The left panel shows the truncation down to 10$\au$, the middle panel shows to 30$\au$, and the right panel shows to 50$\au$. The central stars X-ray luminosity $L_{\rm X} = 10^{30} \rm erg s^{-1}$.}
\label{fig:truncation_times}
\end{figure*}

\subsection{Environments that show biggest difference when including Internal Photoevaporation}

Whilst statistics of disc lifetimes could yield indications into whether discs are viscous or wind driven, the evolution of the size of a disc may still be a useful metric.
The above sections showed that when including internal photoevaporation, weak FUV environments may not be the best environment to find such trends, since the disc evolution for all discs there is quite similar.

Instead, more intermediate environments may be better targets, since it was seen in fig. \ref{fig:r90_IPE} that there are significant differences in their evolution later in the disc lifetime.
To this end, in fig. \ref{fig:truncation_times} we plot the time of truncation to different radii in the disc as a function of $\psi$ and the external environment.
The central star luminosity was again set to $L_{\rm X}=10^{30} \rm erg s^{-1}$.
The left panel shows the truncation time to 10$\au$, with the middle and right panels showing the times to 30 and 50$\au$ respectively.

Looking first at the left panel, it is clear to see that wind driven discs taken longer to truncate down to 10 $\au$ than viscous discs.
This is all apart from the far upper right of the panel, in the most extreme environment where it is slightly shorter.
The horizontal lines that appear in the plots for both viscous and wind driven discs, show the qualitative evolution of the discs changing, where holes are able to form due to internal photoevaporation, and the outer discs are able to be quickly or slowly dispersed.
The trend of wind driven discs taking longer to truncate than viscous discs continues when looking at the middle panel of fig. \ref{fig:truncation_times}, those to 30 $\au$, but the differences here are to a lesser extent.
Additionally for the stronger environments, there appears to be little difference now between the two processes, as external photoevaporation dominates the evolution of the disc in this outer region.
Going now to the right panel of fig. \ref{fig:truncation_times}, showing the truncation time to 50 $\au$, the effect of external photoevaporation is clear for environments $>10^{3} \rm G_0$ where there are minimal differences as a function of $\psi$.
However for the weaker environments, the viscous discs now take longer to truncate to this level, highlighting the early effect of viscous expansion in setting an equilibrium.
The reason this trend does not extend to lower levels, is due to internal photoevaporation later in the disc lifetime opening a hole more quickly in the viscous discs than in the wind driven discs, allowing them to lose their outer disc more quickly, and truncating down at a faster rate.

\begin{figure*}
\centering
\includegraphics[scale=0.39]{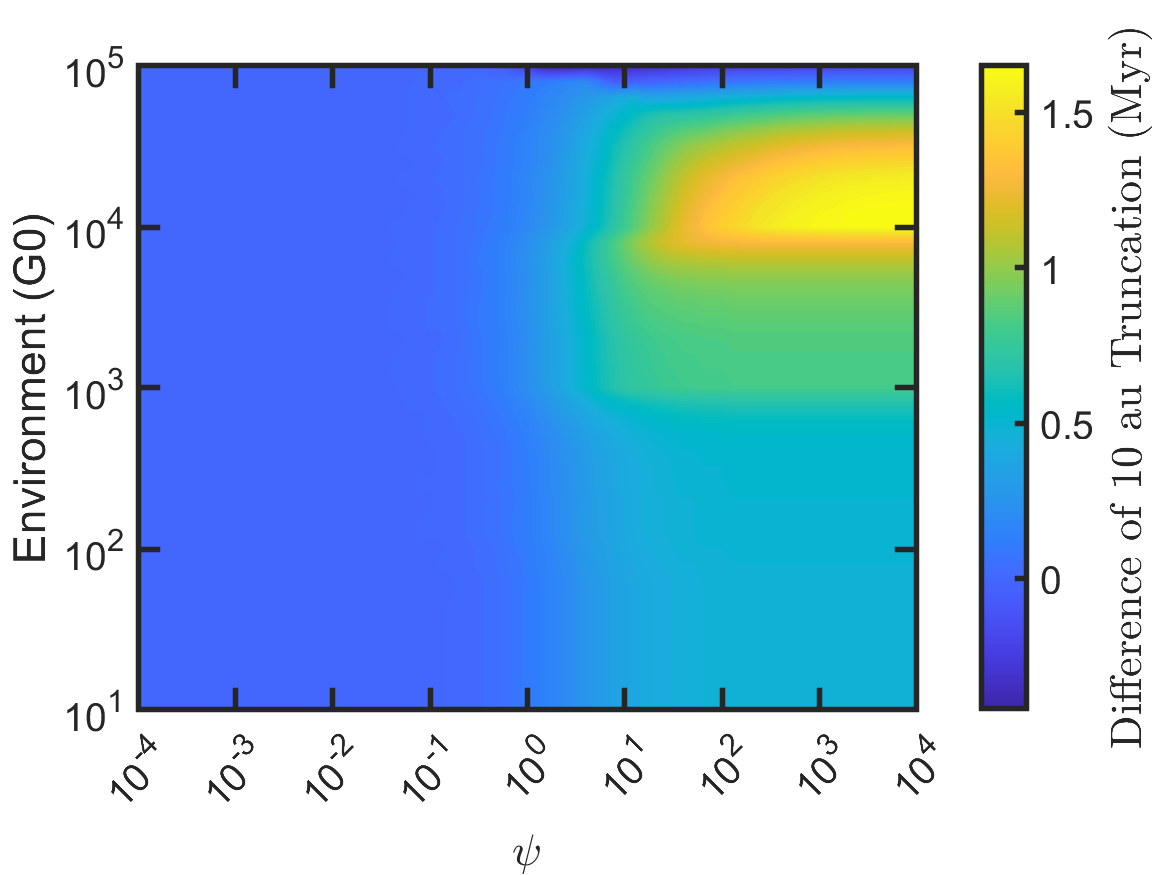}
\includegraphics[scale=0.39]{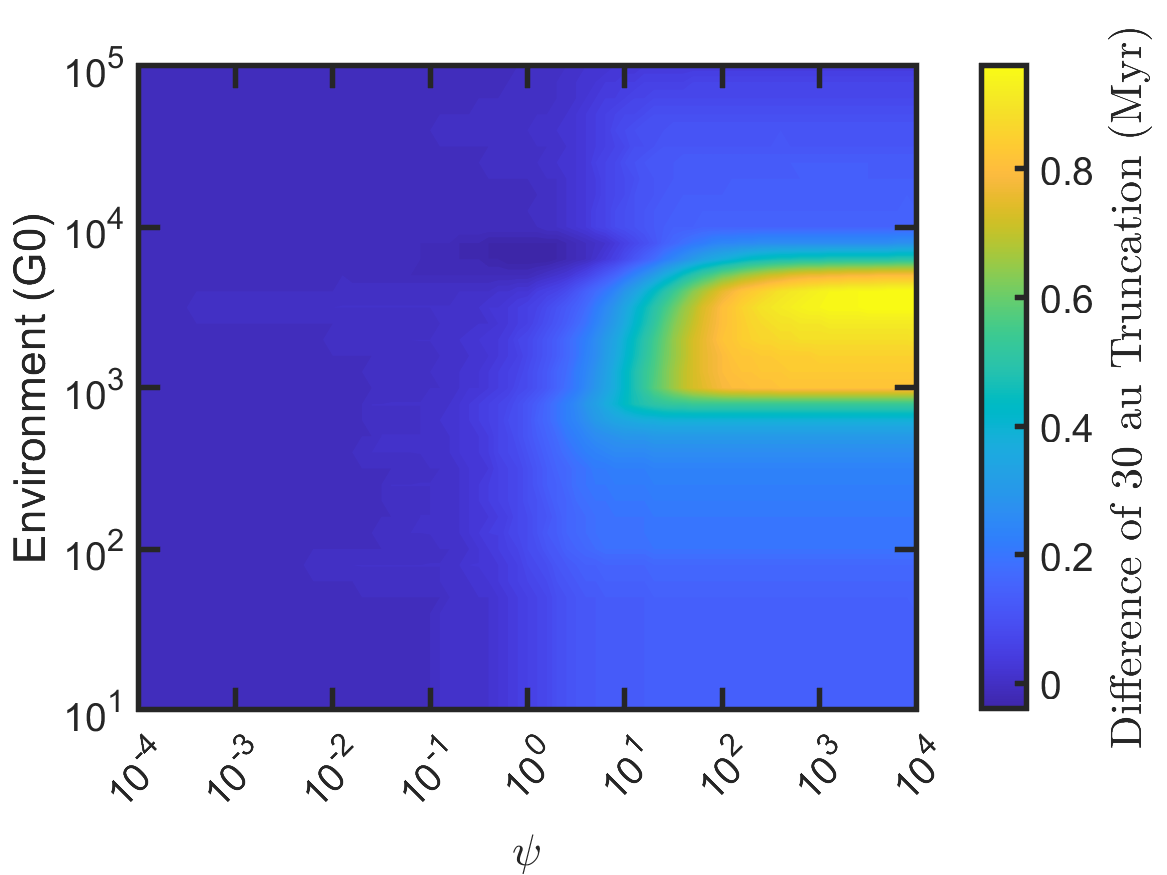}
\includegraphics[scale=0.39]{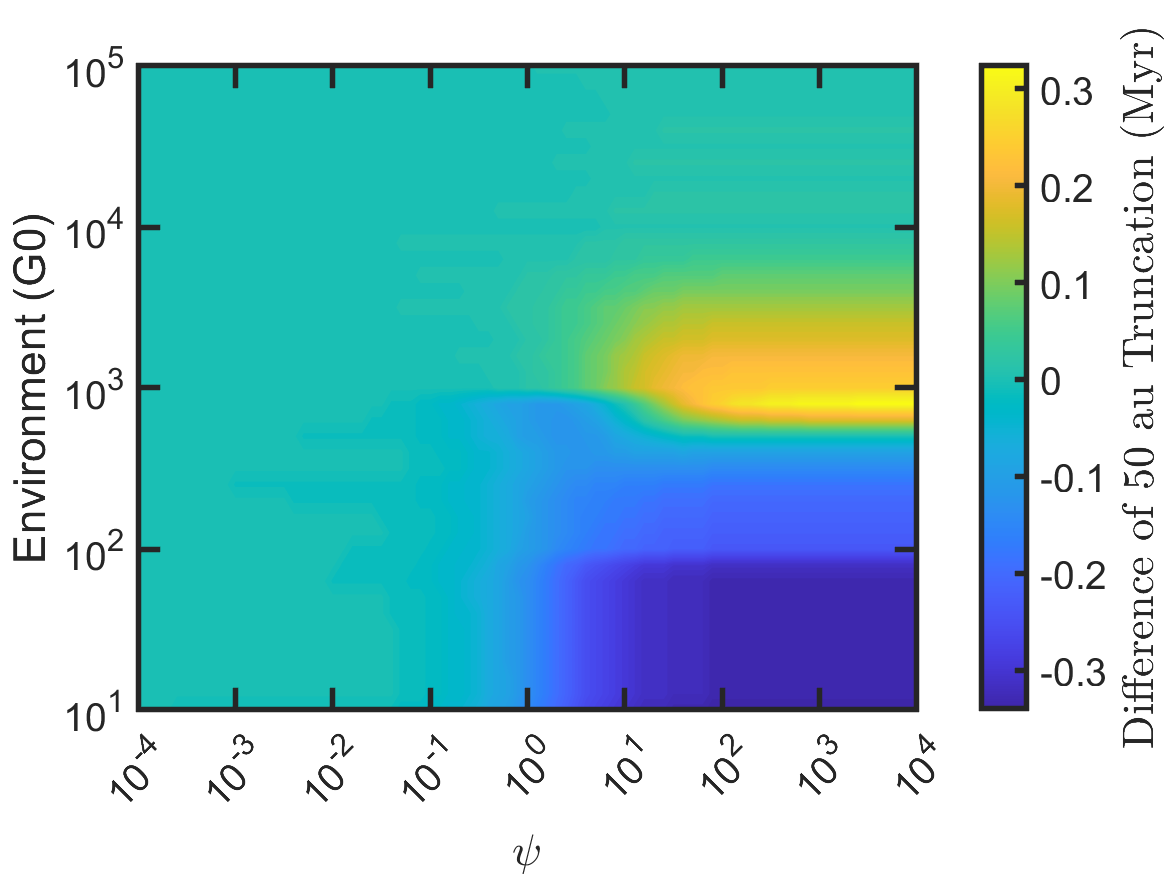}
\caption{Contours plots showing the differences in truncation times to specific radius for discs in different environments and different relative strengths of viscosity and MHD winds, to equivalent purely viscous discs only. The left panel shows the truncation down to 10$\au$, the middle panel shows to 30$\au$, and the right panel shows to 50$\au$.}
\label{fig:truncation_times_diff}
\end{figure*}

Whilst fig. \ref{fig:truncation_times} showed the truncation times, fig. \ref{fig:truncation_times_diff} shows the difference in truncation times to that observed for a viscous disc, i.e. $\psi=10^{-4}$ or the far left of each panel in fig. \ref{fig:truncation_times}.
Interestingly the largest differences seem to be in the intermediate and stronger environments.
Indeed in truncating down to 10$\au$ the larger difference is in environments $\sim10^4 \rm G_0$.
Here the wind driven discs remain larger for much longer than their viscous counterparts, since in these regions it is the viscous expansion that continues to feed stronger photoevaporative winds, whereas the disc winds reduce the supply.
This results in those wind driven discs staying larger for longer periods of time, sometimes up to 1.5 Myr.
This trend is also seen in the middle panel, where the greatest difference is around $\sim5000 \rm G_0$, with again the wind driven discs remaining larger.
As was shown in fig. \ref{fig:r90_IPE}, this is mainly due to internal photoevaporation opening the hole and the outer disc depleting more quickly in the viscous discs.
But nonetheless this feature may be seen in more evolved star forming regions.
Finally, looking at the right panel of fig. \ref{fig:truncation_times_diff}, the same feature appears for truncating down to 50 $\au$, but it is extremely narrow now around $10^3 \rm G_0$.
Additionally the temporal differences are only 0.3 Myr at most, which when taking into account uncertainties in star forming region ages, this becomes difficult to disentangle from other possibilities, i.e. when the stars actually form.

An extra point to take from the truncation time difference to 50$\au$ is that for weak environments, the wind driven discs are seen to truncate at a faster rate, but only by again 0.3 Myr.
In taking all of the truncation time differences into account, it is clear that it is important to account for the role of environment when attempting to determine whether observed discs evolve through viscosity or MHD winds.
It is also important to know the stage of it's lifetime that the disc is in, as this could also aid in analysing discs in stronger environments, where larger differences between viscous and wind driven models seem to exist.

\section{Summary and Conclusions}
\label{sec:conclusions}

In this work we have explored the effects that photoevaporation, both internally and externally driven, have on the evolution of viscous and MHD wind driven protoplanetary discs.
The main aim of this work was to determine whether there were observational differences between viscous and wind driven discs \citep[which has been proposed, e.g. ][]{Tabone22b,Tabone22,Manara23,Alexander23,Somigliana23} when photoevaporation was also considered.
We performed a broad parameter study for external photoevaporation, looking at both weak and strong UV environments, before then including the effects of a weak internal photoevaporative wind.
We draw the following main conclusions from this work.\\

1. Including external photoevaporation halts viscous expansion when the expansion rate equals the external photoevaporative mass loss rate. This occurs even in weak UV environments, differing from what is shown in viscous evolutionary models. When comparing viscous disc models to MHD wind driven disc models, the observable differences, i.e. disc radius, is therefore reduced making it harder to discern whether discs are viscous or wind driven. This highlights the importance of including even weak ($\sim10$\,G$_0$) external photoevaporation when exploring models of protoplanetary disc evolution.\\

2. Whilst there are observable differences between evolving viscous and wind driven disc models in weak UV environments, these differences become negligible in stronger environments, e.g. $>10^3\rm G_0$. This is due to external photoevaporation dominating the mass evolution of the outer disc regions, efficiently truncating the discs down to a compact size. This reduces and ultimately washes outs differences, e.g. in observable disc radii, between viscous and wind driven disc models.\\

3. Analytical models predicted that there would be observable differences in the relations between mass accretion rate, disc mass and disc radius \citep[e.g.][]{Lodato17,Somigliana20,Somigliana22,Somigliana23,Manara23}. Our results here show that whilst we find some of these predictions to be qualitatively similar, significance and observability are reduced. For example, viscous discs typically tended to have larger mass accretion rates for a similar disc mass as wind driven discs. However differences are only by a factor few that could equally be found when altering the initial properties of the disc, including how massive and how compact it is. Additionally the trends in the remaining lifetime of the discs are found to to be similar, irrespective of the mechanism of angular momentum transport in the disc. All of these factors taken together show that the possibility of discerning whether discs are viscous or wind driven is much harder and more degenerate than previously considered.\\

4. The prospects of comparing viscous discs to wind driven discs are further diminished when internal photoevaporation processes are included. Our results show that even a weak internal photoevaporative wind is able to dominate the evolution of protoplanetary discs in low UV environments. It does this by removing the mass in the intermediate regions of the disc ($10<r<100\au$), before opening a hole in the disc, and quickly dispersing the outer disc. This effect occurs before the discs have sufficiently evolved through either viscous or wind driven evolution. With internal photoevaporation dominating, this washes out many previous differences found between viscous and wind driven discs in weak UV environments.\\

5. Whilst discs in weak UV environments provide less insight into the problem of viscous versus MHD wind driven evolution when including internal photoevaporation, more intermediate environments may provide times where there \textit{are} significant differences. This is mainly where internal photoevaporation opens a hole in the disc and the outer disc quickly disperses. Due to viscous expansion feeding external photoevaporative winds, this occurs earlier for viscous discs, and so disc radii at these late times in a discs lifetime can be used to discern between viscous or wind driven discs, assuming the other parameters, e.g. strength of the internal field are adequately constrained. \\

Overall, this work shows the importance of including photoevaporative effects in models of disc evolution. Even when the external environment or the stellar X-ray luminosity are considered weak, there is sufficient impact to affect the evolution of either viscous or wind driven discs, either through halting viscous expansion, or quickly dispersing the disc once a hole opens.

Aspects not considered in this work, but which would affect the comparison of star forming regions and disc evolution is the evolution of the surrounding star forming region itself. Not only do stars and their subsequent discs form at different times, the external radiation field experienced by an evolving disc changes over time. Including the star formation rate within comparisons of disc lifetimes reveals a degeneracy in determining the evolution pathway of discs in different environments \citep{Coleman22}, whilst protoplanetary discs can often be shielded for a sufficient portion of their early lifetime before emerging in to stronger FUV environments \citep{Ali19,Qiao22, 2023MNRAS.520.5331W}. Both of these effects will also be relevant to some degree when using disc properties to determine whether discs are viscously or wind driven. Given that our work shows that the qualitative evolution of protoplanetary discs are similar for viscous and wind driven discs, and the main differences is when such discs go through certain transitional stages in their lifetimes, the added degeneracy of the effects of the star forming region itself further complicates matters.
Ultimately, understanding the fundamental properties of a protoplanetary disc, it's central star, and it's recent history will be important to eliminate some of these degeneracies, in order to address the question of whether they are viscous or wind driven.

Additionally, whilst this work assumed that all disc evolution processes work in tandem, it is still an open question of how do they interact with each other. Signatures of MHD \citep{Campbell-White23}, internal \citep{Rab23} and external \citep[e.g.][]{1994ApJ...436..194O, 2016ApJ...826L..15K, 2017MNRAS.468L.108H, 2021MNRAS.501.3502H} photoevaporative winds have been associated with nearby protoplanetary discs. With evidence for all of these processes occurring, an overlap in space and/or time must exist, and so future work is required to further understand how these processes interact with each other. Indeed, recent work has shown that MHD winds and internally photoevaporative winds do work in tandem, with MHD winds dominating mass loss rates when the magnetic field is high or X-ray luminosity of the star is weak, and vice versa for internal photoevaporative winds \citep{Sarafidou23}. With such an improved understanding of the interplay between all of these processes, the validity of disc evolution models incorporating these processes will be improved and we can begin to fully understand how protoplanetary discs evolve.
Only then, by including more complex models, and in understanding the properties of observed protoplanetary discs better, will it be possible to understand how protoplanetary discs evolve, and then subsequently how might planetary systems form within them.

\section*{Data Availability}
The data underlying this article will be shared on reasonable request to the corresponding author.

\section*{Acknowledgements}
The authors thank the anonymous referee for providing useful and interesting comments that improved the paper.
GALC acknowledges support from STFC through grants ST/P000592/1 and ST/T000341/1.
TJH acknowledges funding by a Royal Society Dorothy Hodgkin Fellowship and UKRI ERC Consolidator Grant guarantee funding (EP/Y024710/1).
This research utilised Queen Mary's Apocrita HPC facility, supported by QMUL Research-IT (http://doi.org/10.5281/zenodo.438045).
This work was performed using the DiRAC Data Intensive service at Leicester, operated by the University of Leicester IT Services, which forms part of the STFC DiRAC HPC Facility (www.dirac.ac.uk). The equipment was funded by BEIS capital funding via STFC capital grants ST/K000373/1 and ST/R002363/1 and STFC DiRAC Operations grant ST/R001014/1. DiRAC is part of the National e-Infrastructure.

\vspace{-0.2cm}
\bibliographystyle{mnras}
\bibliography{references}{}

\vspace{-0.2cm}
\label{lastpage}
\end{document}